\documentclass[12pt, a4paper]{article}
\usepackage{graphicx}
\usepackage{amssymb}
\usepackage{amsmath}
\usepackage{bm}
\usepackage{color}
\usepackage{theorem}
\usepackage{subcaption}
\usepackage{listings}
\usepackage{colortbl}
\usepackage{tabularx}
\usepackage{multirow}
\usepackage{longtable}
\usepackage{enumerate}
\usepackage[utf8]{inputenc}
\usepackage[T1]{fontenc}
\usepackage{lmodern}
\usepackage[top=30truemm,bottom=30truemm,left=25truemm,right=25truemm]{geometry}

\usepackage[sort&compress,numbers, merge]{natbib}

\definecolor{Orange}{cmyk}{0,0.61,0.87,0}
\definecolor{JungleGreen}{cmyk}{0.99,0,0.52,0}
\definecolor{OliveGreen}{cmyk}{0.64,0,0.95,0.40}
\definecolor{Brown}{cmyk}{0,0.81,1,0.60}
\definecolor{RoyalBlue}{cmyk}{0.71,0.53,0,0.12}
\definecolor{Gray}{cmyk}{0,0,0,0.40}
\definecolor{LightPink}{cmyk}{0.0,0.25,0,0}
\definecolor{LLightPink}{cmyk}{0.0,0.10,0,0}
\definecolor{LightBlue}{cmyk}{0.25,0,0,0}
\definecolor{LightGray}{cmyk}{0,0,0,0.2}

\allowdisplaybreaks[1]

\usepackage[colorlinks=true, linkcolor=OliveGreen, citecolor=RoyalBlue,
urlcolor=RoyalBlue]{hyperref}

\renewcommand{\thefootnote}{\fnsymbol{footnote}}

\begin{document}

\begin{titlepage}

  \begin{flushright}
    {\tt
    }
\end{flushright}

\vskip 1.35cm
\begin{center}

{\Large
{\bf
Thermal Leptogenesis in the Minimal Gauged $U(1)_{L_\mu-L_\tau}$ Model
}
}

\vskip 1.5cm

\renewcommand*{\thefootnote}{\fnsymbol{footnote}}
A.~Granelli$^{a,b~}$\footnote{\href{mailto:alessandro.granelli@unibo.it}{\tt alessandro.granelli@unibo.it}},
K.~Hamaguchi$^{c,d~}$\footnote{
\href{mailto:hama@hep-th.phys.s.u-tokyo.ac.jp}{\tt
 hama@hep-th.phys.s.u-tokyo.ac.jp}},
N.~Nagata$^{c~}$\footnote{\href{mailto:natsumi@hep-th.phys.s.u-tokyo.ac.jp}{\tt natsumi@hep-th.phys.s.u-tokyo.ac.jp}},
M.~E.~Ramirez-Quezada$^{c,e~}$\footnote{\href{mailto:me.quezada@hep-th.phys.s.u-tokyo.ac.jp}{\tt me.quezada@hep-th.phys.s.u-tokyo.ac.jp}} 
and J.~Wada$^{c~}$\footnote{\href{mailto:wada@hep-th.phys.s.u-tokyo.ac.jp }{\tt wada@hep-th.phys.s.u-tokyo.ac.jp}}

\vskip 0.8cm
{\it $^{a}$Dipartimento di Fisica e Astronomia, Università di Bologna, via Irnerio 46, 40126,}\\[2pt]
{\it $^{b}$INFN, Sezione di Bologna, viale Berti Pichat 6/2, 40127, Bologna, Italy,}\\[2pt]
{\it ${}^c$Department of Physics, University of Tokyo, Bunkyo-ku, Tokyo 113--0033, Japan,} \\[2pt]
 {\it ${}^d$Kavli Institute for the Physics and Mathematics of the Universe (Kavli IPMU), University of Tokyo, Kashiwa 277--8583, Japan,}\\[2pt]
  {\it $^e$Dual CP Institute of High Energy Physics, C.P. 28045, Colima, M\'exico.}

\date{\today}

\vskip 1.5cm

\begin{abstract}
We discuss the thermal leptogenesis mechanism within the minimal gauged U(1)$_{L_\mu-L_\tau}$ model to explain the observed baryon asymmetry of the Universe (BAU).  
In such framework, the phases of the Pontecorvo–Maki–Nakagawa–Sakata neutrino mixing matrix and the sum of the Standard Model neutrino masses are predictable because of a restricted neutrino mass matrix structure. Additionally, in the context of thermal leptogenesis, the BAU can be computed in terms of the three remaining free variables that parameterise the right-handed neutrino masses and their Yukawa couplings to the Higgs and lepton doublets. We identify the ranges of such parameters for which the correct BAU can be reproduced. We adopt the formalism of the density matrix equations to fully account for flavour effects and consider the decays of all the three right-handed neutrinos. Our analysis reveals that thermal leptogenesis is feasible within a wide parameter space, specifically for Yukawa couplings ranging from approximate unity to $\mathcal{O}(0.03-0.05)$ and mass of the lightest right-handed neutrino $M_1\gtrsim 10^{11-12}\,\text{GeV}$, setting a leptogenesis scale in the considered model which is higher than that of the non-thermal scenario.
\end{abstract}

\end{center}
\end{titlepage}
\renewcommand{\thefootnote}{\arabic{footnote}}
\setcounter{footnote}{0}

\section{Introduction}
\label{sec:intro}

There is astrophysical and cosmological evidence for the existence of a matter-antimatter asymmetry in the present Universe. The baryon-to-photon ratio parameterises the baryon asymmetry of the Universe (BAU), 
\begin{equation}
\eta_B = \frac{n_B - n_{\bar{B}}}{n_\gamma},
\end{equation}
where $n_B$, $n_{\bar{B}}$, and $n_\gamma$ are the number densities of baryons, anti-baryons, and photons, respectively. The baryon-to-photon ratio has been determined independently from observations of the Cosmic Microwave Background (CMB) anisotropies and Big Bang Nucleosynthesis (BBN) estimates. Both estimations are consistent with the best-fit value $\eta_B\simeq 6.1\times 10^{-10}$~\cite{Planck:2018vyg, Cooke_2018}.  One compelling mechanism that explains the observed BAU is leptogenesis \cite{Fukugita:1986hr}
based on the existence of right-handed neutrinos and their out-of-equilibrium decays in the early Universe. In the simplest thermal leptogenesis scenario, the CP-violating, out-of-equilibrium decays of the right-handed neutrinos generate a lepton asymmetry, which is converted into a baryon asymmetry by the sphaleron processes predicted by the SM~\cite{Kuzmin:1985mm}. The leptogenesis mechanism for the BAU generation can be studied in a wide range of models beyond the SM, providing a valuable window into the study of new physics (see, e.g., Ref.~\cite{Bodeker:2020ghk} for a recent review on the topic and references therein).

In this paper, we focus on a model that exhibits a novel anomaly-free U(1) gauge symmetry denoted by U(1)$_{L_\mu - L_\tau}$\cite{Foot:1990mn,He:1990pn,PhysRevD.44.2118,Foot:1994vd}, where $L_e$, $L_\mu$ and $L_\tau$ stand for the lepton number (flavour) associated to the electron ($e$), muon ($\mu$) and tauon ($\tau$), respectively. The model is implemented within the framework of the type-I seesaw mechanism \cite{Minkowski:1977sc, Yanagida:1979as, Gell-Mann:1979vob, Glashow:1979nm, Mohapatra:1979ia}, hence providing an explanation for the smallness of SM neutrino masses through the introduction of three heavy right-handed neutrinos $N_{1,2,3}$ with masses $M_{1,\,2,\,3}$.
The  symmetry of the model imposes constraints on the neutrino mass structure, since the second and third-generation leptons, $\mu$ and $\tau$, are charged under U(1)$_{L_\mu-L_\tau}$ while the electron is not. Consequently, the Dirac mass matrix has a simple  diagonal structure, and only certain components of the Majorana matrix are non-zero. This simple mass structure is insufficient to explain neutrino oscillation data~\cite{Branco:1988ex, Choubey:2004hn}, and hence, the U(1)$_{L_\mu-L_\tau}$ must be broken. This is typically achieved by introducing a scalar field that has non-zero U(1)$_{L_\mu-L_\tau}$ charge and a vacuum expectation value (VEV), which breaks the gauge symmetry giving mass to the U(1)$_{L_\mu-L_\tau}$  gauge bosons. We refer to this model as the ``minimal gauged U(1)$_{L_\mu-L_\tau}$ model''. 
The model has a strong predictive power \cite{Branco:1988ex, Choubey:2004hn, Araki:2012ip, Heeck:2014sna, Crivellin:2015lwa,Plestid:2016esp,Asai:2017ryy, Asai:2018ocx, Asai:2019ciz,Asai:2020qax} because of the so-called two-zero minor structure \cite{Araki:2012ip, Heeck:2014sna, Crivellin:2015lwa, Asai:2017ryy, Asai:2018ocx, Asai:2019ciz} of the neutrino mass matrix.

The aim of the present work is to study thermal leptogenesis in the minimal gauged U(1)$_{L_\mu-L_\tau}$ model.\footnote{For a study of the non-thermal leptogenesis mechanism within the minimal gauged U(1)$_{L_\mu-L_\tau}$ model, see Ref.~\cite{Asai:2020qax}. Leptogenesis in (non-minimal) gauged U(1)$_{L_\mu-L_\tau}$ models is also discussed in Refs.~\cite{Borah:2021mri, Qi:2022kgs, Eijima:2023yiw}.} 
Since the neutrino mass structure is highly restricted in the model, it is not obvious whether the observed BAU is obtainable through thermal leptogenesis. 
In a
widely-considered setup for thermal leptogenesis \cite{Blanchet:2012bk}, the right-handed neutrinos mass spectrum is strongly hierarchical, $M_1 \ll M_2 \ll M_3$, and the lepton asymmetries in the three flavours evolve equally. 
This scenario has been extensively studied solving the unflavoured Boltzmann Equations (BEs) and is subject to a model-independent bound on the mass scale of leptogenesis, reading $M_1\gtrsim 10^9\,\text{GeV}$, below which the requisite CP-asymmetry is too small to get the observed value of the BAU \cite{Davidson:2002qv, Buchmuller:2002rq, Ellis:2002xg, Buchmuller:2003gz, BUCHMULLER2005305}. However, in our model, the condition of a strong mass hierarchy is not satisfied in many parts of the parameter space, and we expect the decay of each of the three right-handed neutrinos to contribute to the generation of the BAU.

Charged lepton flavour effects can also play a crucial role in the generation of the BAU
\cite{Barbieri:1999ma, Nielsen:2002pc,Endoh:2003mz, Nardi:2006fx,Abada:2006fw,Abada:2006ea, Simone_2007, Blanchet_2007, Blanchet_2013, Dev:2017trv}. 
Specifically, the unflavoured scenario  is valid only in the single-flavour regime for temperatures above $\sim 10^{12}\,\text{GeV}$, when the processes mediated by the charged lepton SM Yukawa couplings are out-of-equilibrium.  In the two-flavour regime $10^{9}\ll T/\text{GeV}\ll 10^{12}$, processes induced by the $\tau$-Yukawa coupling occur at a rate $\Gamma_\tau$ much larger than the Hubble expansion rate $H$, which indicates that these processes are in thermal equilibrium. Consequently, the asymmetries in the lepton charge $L_\tau$ and $L_\mu + L_e$ evolve differently in this regime. Similarly, in the three-flavour regime for $T\ll 10^{9}\,\text{GeV}$, processes mediated by the $\mu$-Yukawa coupling with rate $\Gamma_\mu$ are also in thermal equilibrium ($\Gamma_\mu \gg H$), leading to the individual evolution of $L_e$, $L_\mu$ and $L_\tau$. Given that the mass scales of interest cover different flavour regimes, the simplest unflavoured scenario is not applicable to our study, and we have to consider the impact of charged lepton flavour effects on the BAU generation.

The study of thermal leptogenesis in the (three-) two-flavour regime can be conducted using the formalism of (three-) two-flavoured BEs, provided that the processes mediated by the ($\mu$- and $\tau$-) $\tau$-Yukawa couplings are sufficiently fast. In this formalism, the equations for the asymmetries in  ($L_\tau$, $L_\mu$ and $L_e$) $L_\tau$ and $L_\mu + L_e$  are different and solved separately. However, to accurately account for flavour effects, it is more precise  to trace the evolution of the elements of the density matrix of the lepton flavour system with the Density Matrix Equations (DMEs) \cite{Simone_2007, Blanchet_2007,Blanchet_2013, Dev:2017trv}. This approach is particularly useful when the processes mediated by the charged lepton Yukawa couplings are neither infinitely fast nor their effects negligible, such as at the transitions between different flavour regimes, and has been shown to lead to different predictions with respect to the BEs \cite{Simone_2007, Blanchet_2007,Blanchet_2013, Dev:2017trv, Moffat:2018wke, Granelli:2021fyc}.

The presence of multiple decaying right-handed neutrinos also has implications for the generation of the BAU, specifically because of the effects of heavy neutrino flavours. The right-handed neutrinos couple to different superpositions of flavour states, whose interactions can induce additional decoherence effects in the context of the DMEs \cite{Blanchet_2013} (see also Ref.~\cite{Dev:2017trv}). These effects can be particularly relevant when the right-handed neutrinos do not have a strongly hierarchical mass spectrum, which is the case in certain parts of the parameter space of the considered model where $M_2\lesssim 3M_1$ and $M_3\lesssim 3M_2$. In this regime, the different superposition of flavour states, associated with the different right-handed neutrinos, are simultaneously present in the Universe.

In this work, we consider the formalism of the DMEs with three decaying right-handed neutrinos to fully account for all the relevant effects mentioned above. By performing a numerical scan of the parameter space, we investigate for which values of the parameters of the minimal gauge U(1)$_{L_\mu-L_\tau}$ model the DMEs are successful in reproducing the observed BAU. We solve numerically the DMEs for thermal leptogenesis using the Python package \texttt{ULYSSES} \cite{Granelli:2020pim, Granelli:2023vcm}, that is a freely accessible code for the numerical evaluation of the BAU in the context of leptogenesis. Its major features are the variety of equations available, allowing for comparisons between the BEs and DMEs in the different regimes, and a rapid evaluation, making scans of large parameter spaces feasible over relatively short periods of time. 

The paper is structured as follows. In Sec.~\ref{sec:model}, we present the minimal gauged U(1)$_{L_\mu-L_\tau}$ model and analyse the corresponding neutrino mass structure. In Sec.~\ref{sec:LG}, we discuss in more details the mechanism of thermal leptogenesis and the formalism of DMEs. We introduce the DMEs with three decaying right-handed neutrinos and the corresponding CP-asymmetry parameters, which are crucial for understanding the generation of the lepton and baryon asymmetries in the early Universe. The results of the scan of the parameter space for viable leptogenesis is presented in Sec.~\ref{sec:results}, and we finally conclude in Sec.~\ref{sec:conclusions}.

\section{The Minimal Gauged \texorpdfstring{U(1)$_{L_\mu-L_\tau}$}{U} Model}
\label{sec:model}

In the minimal gauged U(1)$_{L_\mu-L_\tau}$ model, the second and third-generation leptons, namely those of lepton flavour $\mu$ and $\tau$, carry charges of $+1$ and $-1$, respectively. Notably, the first-generation leptons (of lepton flavour $e$), quarks, and the Higgs field in the Standard Model (SM) are not charged under this particular gauge symmetry.
Moreover, we introduce three right-handed sterile neutrinos, $N_e$, $N_\mu$, and $N_\tau$, each described by a Weyl spinor that transforms under the $(0,\,\frac{1}{2})$ representation of the Lorentz group (thus right-handed), which are singlets under the SM gauge group (thus sterile) and carry the U(1)$_{L_\mu - L_\tau}$ charges 0, $+1$, and $-1$, respectively. 
Additionally, we introduce a scalar boson $\sigma$, which is a singlet under the SM gauge group and carries the U(1)$_{L_\mu - L_\tau}$ charge $+1$. This scalar field develops a VEV that spontaneously breaks the U(1)$_{L_\mu - L_\tau}$ gauge symmetry. We summarise the charges of the field content of this model in Table~\ref{tab:model}, denoting the left-handed SU(2) lepton doublets of flavour $\alpha$ with $\ell_{\alpha}$ and the right-handed charged leptons with $\alpha_R$, $\alpha = e,\,\mu,\,\tau$. We adopt a two-component spinor notation (see, e.g., Ref.~\cite{Dreiner:2008tw}) as in Ref.~\cite{Asai:2017ryy}.
\newcolumntype{C}{@{}>{\centering\arraybackslash}X}
\begin{table}[h!]
 \begin{tabularx}{\textwidth}{c|C|C|C|C|c}
\hline
\rowcolor[gray]{.95}
    \multicolumn{6}{c}{\bf $L_\mu-L_\tau$ Charges of the Field Content}\\
    \hline
    \hline
  & $\ell_{e}, e_R, N_e$ & $\ell_{\mu}, \mu_R, N_\mu$ & $\ell_{\tau }, \tau_R, N_\tau$ &  ~~~~$\sigma$ ~~~~& Others  \\
 \hline
 ${L_\mu - L_\tau}$& $0$ &  $+1$ & $-1$ & $+1$ & $0$  \\
 \hline
 \hline
 \end{tabularx}
 \caption{The ${L_\mu-L_\tau}$ charges of the field content in the considered minimal gauged U(1)$_{L_\mu-L_\tau}$ model.}
   \label{tab:model}
 \end{table}

For the purposes of our discussion, we focus on the new leptonic interactions involving the right-handed neutrinos and their Majorana mass terms, reading 
\begin{align}
 \Delta {\cal L} = 
&-\lambda_e N_e^c (\ell_{e} \cdot \Phi)
-\lambda_\mu N_\mu^c (\ell_{\mu} \cdot \Phi)
-\lambda_\tau N_\tau^c (\ell_{\tau} \cdot \Phi) \nonumber \\
&-\frac{1}{2}M_{ee} N_e^c N_e^c 
- M_{\mu \tau} N_\mu^c N_\tau^c 
- \lambda_{e\mu} \sigma N_e^c N_\mu^c
- \lambda_{e\tau} \sigma^* N_e^c N_\tau^c +\text{h.c.} ~,
\label{eq:LagMinMod}
\end{align}
where the dots indicate the contraction of the SU(2) indices between the lepton doublets and the Higgs doublet $\Phi$. Additionally,  $(N_\alpha^c)_a\equiv \varepsilon_{ab} (N_\alpha^*)_b$, where $\alpha = e,\,\mu,\,\tau$ and $\varepsilon_{ab}$ is the antisymmetric tensor of the spinor indices $a,\,b$. The interaction terms in Eq.~\eqref{eq:LagMinMod} lead to neutrino mass terms after the Higgs field $\Phi$ and singlet scalar $\sigma$ acquire their VEVs, denoted as $\langle \Phi \rangle = v/\sqrt{2}$ and $\langle \sigma \rangle$, respectively. These mass terms can be expressed as follows:
\begin{align}
 {\cal L}_{\rm mass} &= -(\nu_{e}, \nu_{\mu}, \nu_{\tau}) {\cal M}_D 
\begin{pmatrix}
N_e^c\\ N_\mu^c\\ N_\tau^c 
\end{pmatrix}
- \frac{1}{2}(N_e^c, N_\mu^c, N_\tau^c) {\cal M}_R 
\begin{pmatrix}
 N_e^c \\ N_\mu^c \\ N_\tau^c 
\end{pmatrix}
+\text{h.c.} ~,
\end{align}
where $\nu_\alpha$, are the left-handed Weyl spinors describing the SM neutrinos of lepton flavour $\alpha$, ${\cal M}_D$ is the Dirac mass matrix and $ {\cal M}_R$ is the Majorana mass matrix given by,
\begin{equation}
 {\cal M}_D = \frac{v}{\sqrt{2}}
\begin{pmatrix}
 \lambda_e & 0& 0\\
 0 & \lambda_\mu & 0 \\
 0 & 0 & \lambda_\tau 
\end{pmatrix}
~,\qquad
{\cal M}_R =
\begin{pmatrix}
 M_{ee} & \lambda_{e\mu} \langle \sigma \rangle & \lambda_{e\tau} 
\langle \sigma \rangle \\
 \lambda_{e\mu} \langle \sigma \rangle & 0 & M_{\mu\tau} \\
\lambda_{e\tau} \langle \sigma \rangle & M_{\mu\tau} & 0
\end{pmatrix}
~,
\end{equation}
respectively. Notably the Dirac mass matrix $\mathcal{M}_D$ is diagonal, while the $(\mu,\mu)$ and $(\tau, \tau)$ components in the Majorana mass matrix $\mathcal{M}_R$ are zero due to the U(1)$_{L_\mu - L_\tau}$ gauge symmetry. 
This particular structure leads to interesting predictions for neutrino observables~\cite{Lavoura:2004tu, Lashin:2007dm, Crivellin:2015lwa, Asai:2017ryy, Asai:2018ocx}, as discussed below. For the remainder of this discussion, we assume that the Dirac Yukawa couplings $\lambda_e$, $\lambda_\mu$, and $\lambda_\tau$, as well as the VEVs $v$ and $\langle \sigma \rangle$ are real without loss of generality, with $v = 246\,\text{GeV}$; this can always be realised via field redefinition. 

The seesaw master formula can naturally explain the smallness of SM neutrino masses by assuming that the non-zero components in the Majorana mass matrix are much larger than those in the Dirac matrix. The formula relates the light neutrino masses to the masses of the heavy right-handed neutrinos, and their couplings to the SM particles \cite{Minkowski:1977sc, Yanagida:1979as, Gell-Mann:1979vob, Glashow:1979nm, Mohapatra:1979ia},
\begin{equation}
 {\cal M}_{\nu_L} \simeq - {\cal M}_D {\cal M}_R^{-1} {\cal M}_D^T ~.
\label{eq:mnul}
\end{equation}
The mass matrix can be diagonalised, allowing us to express the flavour eigenstates of the neutrinos as linear combinations of the mass eigenstates, 
 \begin{equation}
 U^T {\cal M}_{\nu_L} U =\text{diag}(m_1, m_2, m_3) ~.
 \label{eq:mnuldiag}
\end{equation}
Here, $U$ is  the Pontecorvo-Maki-Nakagawa-Sakata (PMNS) mixing matrix $U$~\cite{Pontecorvo:1967fh, Pontecorvo:1957cp, Pontecorvo:1957qd, Maki:1962mu} and can be parameterised as \cite{Tanabashi:2018oca} 
\begin{equation}
 U = 
\begin{pmatrix}
 c_{12} c_{13} & s_{12} c_{13} & s_{13} e^{-i\delta} \\
 -s_{12} c_{23} -c_{12} s_{23} s_{13} e^{i\delta}
& c_{12} c_{23} -s_{12} s_{23} s_{13} e^{i\delta}
& s_{23} c_{13}\\
s_{12} s_{23} -c_{12} c_{23} s_{13} e^{i\delta}
& -c_{12} s_{23} -s_{12} c_{23} s_{13} e^{i\delta}
& c_{23} c_{13}
\end{pmatrix}\times
\begin{pmatrix}
 1 & 0&0 \\
 0& e^{i\frac{\alpha_{2}}{2}} & 0\\
 0& 0& e^{i\frac{\alpha_{3}}{2}}
\end{pmatrix},
\end{equation}
where $c_{ij} \equiv \cos \theta_{ij}$ and $s_{ij} \equiv \sin \theta_{ij}$ and $\delta,\,\alpha_{2},\,\alpha_3 \in [0, 2\pi]$. The Normal Ordering (NO) of neutrino masses, where $m_1<m_2<m_3$, is the only mass hierarchy that is consistent with neutrino oscillation data in this model, as demonstrated in Ref.~\cite{Asai:2017ryy}. 

Using Eqs.~\eqref{eq:mnul} and \eqref{eq:mnuldiag}, it is possible to express the Majorana mass matrix $\mathcal{M}_R$ as 
\begin{equation}
  {\cal M}_R = - {\cal M}_D^T \, U \, \text{diag}(m_1^{-1}, m_2^{-1}, m_3^{-1})\,  U^T {\cal M}_D ~.
  \label{eq:mr}
\end{equation}
The $(\mu, \mu)$ and $(\tau, \tau)$ components of the right-hand side of the above equation vanish. This is a direct consequence of the two-zero-minor structure~\cite{Lavoura:2004tu, Lashin:2007dm} of ${\cal M}_{\nu_L}$ in this model.\footnote{
Note that this structure is stable against the renormalisation-group effects~\cite{Asai:2017ryy}. 
} 
The vanishing conditions on complex quantities result in four real parameter equations. These equations allow us to predict the values of the lightest neutrino mass $m_1$, as well as the Dirac and Majorana CP phases $\delta$, $\alpha_2$, and $\alpha_3$, as functions of the other neutrino oscillation parameters. These parameters include the neutrino mixing angles $\theta_{12}$, $\theta_{23}$, $\theta_{13}$, and the squared mass differences $\Delta m_{21}^2 \equiv m_2^2 - m_1^2$, and $\Delta m_{31}^2 \equiv m_3^2 - m_1^2$. For each set of input parameters, there exist two possible sets of predictions, as shown in Ref.~\cite{Asai:2017ryy}. Specifically, if the set $(m_1,\, \delta,\, \alpha_2,\, \alpha_3)$ satisfies the two-vanishing conditions, then the set $(m_1,\, 2\pi-\delta,\, 2\pi-\alpha_2,\, 2\pi-\alpha_3)$ also satisfies them. 

For our numerical study, we fix the neutrino mixing angles and the squared mass differences following the \texttt{NuFit} analyses \cite{nufit}, which are comprehensive global fits that include data from all the relevant neutrino oscillation experiments.
The most recent \texttt{NuFit} analyses are conducted both with and without incorporating data from the Super-Kamiokande (SK) experiment. These two approaches yield somewhat different values for the neutrino mixing angles and squared mass differences. 
We summarise the results obtained in the \texttt{NuFit 5.2} analysis \cite{nufit, Esteban:2020cvm} for the best-fit values, $1\sigma$ deviations and $3\sigma$ ranges of the neutrino mixing angles and mass squared differences in Table \ref{Tab::BestFit}. 
\begin{table}[h!]
\centering
\begin{tabularx}{\textwidth}{@{}C|CCCCC@{}}
\hline
\rowcolor[gray]{.95}
    \multicolumn{6}{c}{\bf Neutrino Masses and Mixing Parameters}\\
    \hline
    \hline
        {\small\bf Parameters} & $\theta_{12}$   & $\theta_{13}$  &$\theta_{23}$ &
        $\Delta m_{21}^2$ & $\Delta m_{31}^2$ \\ 
        {\small\bf (units)}&(${}^\circ$) &(${}^\circ$) &(${}^\circ$) & ($10^{-5}\text{ eV}^2$) & ($10^{-3}\text{ eV}^2$) \vspace{.1em}\\ 
        \hline
 \textbf{\footnotesize With SK} & $33.41^{+0.75}_{-0.72}$ & $8.58^{+0.11}_{-0.11}$ & $42.2^{+1.1}_{-0.9}$ & $7.41^{+0.21}_{-0.20}$ & $2.507^{+0.026}_{-0.027}$ \vspace{.1em} \\
 \textbf{\footnotesize $3\sigma$ range} & $[31.31, 35.74]$ & $[8.23, 8.91]$ & $[39.7, 51.0]$ & $[6.82, 8.03]$ & $[2.427,2.590]$ \vspace{.1em} \\
        \hline
 \textbf{\footnotesize Without SK}&  $33.41^{+0.75}_{-0.72}$ & $8.54^{+0.11}_{-0.12}$ & $49.1^{+1.0}_{-1.3}$ & $7.41^{+0.21}_{-0.20}$ & $2.511^{+0.028}_{-0.027}$ \vspace{.1em} \\
  \textbf{\footnotesize $3\sigma$ range} & $[31.31, 35.74]$ & $[8.19, 8.89]$ & $[39.6, 51.9]$ & $[6.82, 8.03]$ & $[2.427,2.590]$ \vspace{.1em} \\
 \hline
 \hline
\end{tabularx}
\caption{
 Best-fit values, $1\sigma$ deviations and $3\sigma$ allowed ranges of the neutrino mixing angles $\theta_{12}$, $\theta_{13}$, $\theta_{23}$, and of the squared mass differences $\Delta m^2_{21}$ and $\Delta m^2_{31}$ in the case of NO light neutrino mass spectrum, from the latest \texttt{NuFit 5.2} analysis \cite{nufit, Esteban:2020cvm}. The two lines of values correspond to the fit with (top) and without (bottom) the inclusion of SK data.
 }
\label{Tab::BestFit}
\end{table}

We observe that, while the inclusion of the SK data substantially affects the best-fit $\pm 1 \sigma$ values of $\theta_{23}$, it does not change dramatically its $3\sigma$ allowed range (the maximal value without SK data is larger by $0.9^\circ$) and the results for $\theta_{12}$, $\theta_{13}$, $\Delta m_{21}^2 $, and $\Delta m_{31}^2$. 
Specifically, the best fit $\pm 1\sigma$ values for $\theta_{23}$ lie below (above) $\pi/4$ with (without) the inclusion of SK data. 
Since the minimal gauged U(1)$_{L_\mu-L_\tau}$ model predictions are highly dependent on the value of $\theta_{23}$, as the sum of neutrino masses diverges for $\theta_{23}=\pi/4$ \cite{Asai:2017ryy}, we examine the cases with and without the inclusion of SK data separately.\footnote{See Fig.~1(b) in Ref.~\cite{Asai:2020qax}, in which the sum of neutrino masses as a function of $\theta_{23}$ is shown. For such figure, the authors of Ref.~\cite{Asai:2020qax} adopted an older version of the neutrino oscillation data (\texttt{NuFit 4.1}), but the behaviour of the same function in our case remains basically the same.} In both cases, we set $\theta_{12}$, $\theta_{13}$, $\Delta m_{21}^2 $, and $\Delta m_{31}^2$ to their best-fit values, while we treat $\theta_{23}$ differently given the implications it has on the sum of neutrino masses predicted by the considered model.

The sum of the neutrino masses is constrained by various cosmological and astrophysical measurements, yielding an upper bound of $\sum_i m_i < (0.12-0.69)\,\text{eV}$ (95$\%$ C.L.), which depends on the adopted model and level of statistical complexity~\cite{LVinZyla:2020zbs, Capozzi_2020} (see also Refs.~\cite{Vagnozzi2017, Planck:2018vyg,RoyChoudhury:2019hls,Ivanov:2019hqk, DES:2021wwk, Tanseri2022}). By adopting the best-fit values for $\theta_{23}$ in the two cases with and without SK data, we obtain $\sum_i m_i\simeq 0.241\,\text{eV}$ and $0.173\,\text{eV}$, respectively. However, in order to evade the aforementioned limitations, some level of complexity and assumptions is required.
We instead set $\theta_{23}$ to its $-\,(+)\,3\sigma$ minimal (maximal) limit for the case with (without) SK data; specifically,  $\theta_{23} = 39.7^\circ\,(51.9^\circ)$, with the other input parameters at their best-fit values. This yields $ 
\sum_i m_i = 0.142\,(0.117)~\mathrm{eV}$, which minimises the sum of neutrino masses and reduces the tension with the cosmological bounds. 
To summarise, in our numerical analysis, we consider the following two sets of input parameters:
\begin{center}
\begin{tabular}{ll}
{\bf Set I} & {\bf Set II} \\
$\theta_{12} = 33.41^\circ$ & $\theta_{12} = 33.41^\circ$ \\
$\theta_{13} = 8.58^\circ$ & $\theta_{13} = 8.54^\circ$ \\
$\theta_{23} = 39.7^\circ$ & $\theta_{23} = 51.9^\circ$ \\
$\Delta m^2_{21} = 7.41\times10^{-5}\,\text{eV}^{2}$&$\Delta m^2_{21} = 7.41\times10^{-5}$\\
$\Delta m^2_{31} = 2.507\times10^{-3}\,\text{eV}^{2}$&$\Delta m^2_{31} = 2.511\times10^{-3}\,\text{eV}^{2}$.
\end{tabular}
\end{center}
The minimal gauged U(1)$_{L_\mu-L_\tau}$ model, when combined with the neutrino mixing angles and squared mass differences in Set I (II), predicts a value of  $m_1 = 0.039\,(0.029)\,\text{eV}$.\footnote{The model also makes predictions about the effective Majorana mass $\langle m_{\beta\beta}\rangle$ which determines the rate of the neutrinoless double-beta decay. The parameter set I (II) predicts $\langle m_{\beta\beta}\rangle\simeq 0.025$ eV (0.016 eV), which is below the current constraint given by the KamLAND-Zen experiment, $\langle m_{\beta\beta}\rangle < 0.036$-0.156 eV~\cite{KamLAND-Zen:2022tow}, and may be probed by future experiments with sensitivities of $\langle m_{\beta\beta}\rangle\simeq {\mathcal O}(0.01)\,\text{eV}$~\cite{Agostini:2017jim,Adams:2022jwx}.
} 
In addition, the PMNS phases are determined to be $\delta \simeq 301^\circ\,(228^\circ)$, $\alpha_2 = 116^\circ\,(225^\circ)$ and $\alpha_3 = 269^\circ\,(70^\circ)$, or equivalently, $\delta = 59^\circ\,(132^\circ)$, $\alpha_2 = 244^\circ\,(135^\circ)$ and $\alpha_3 = 91^\circ\,(290^\circ)$.  The Dirac phase $\delta$ has also been estimated by the \texttt{NuFit} analysis, which reports a $3\sigma$ range of $[144^\circ,350^\circ]$ when including SK data, and $[0,44^\circ]\cup [108^\circ,360^\circ]$ when not including it. Consequently, the set of parameters of Set I with $\delta = 59^\circ$ is disfavoured for more than $3\sigma$.

By specifying the three additional input parameters in the Dirac mass matrix, $\mathcal{M}_D=(v/\sqrt{2})\rm{diag}(\lambda_e, \lambda_\mu, \lambda_\tau)$, it is possible to obtain the right-handed neutrino mass matrix $\mathcal{M}_R$. 
These Yukawa couplings are parameterised, according to Ref.~\cite{Asai:2017ryy}, as
\begin{equation}
  (\lambda_e, \lambda_\mu, \lambda_\tau) = \lambda (\cos \theta, \sin \theta \cos \phi, \sin \theta \sin \phi),
  \label{eq:lthetaphi}
\end{equation}
where we consider the range of $0 \leq \theta\,\rm{and}\, \phi \leq \pi/2$. 
We also limit the value of $\lambda$ to $\lambda \lesssim 1$ to ensure that the Yukawa couplings remain perturbative.  The masses of the right-handed neutrinos are obtained by performing the Takagi diagonalisation on the complex symmetric matrix ${\cal M}_R$
\begin{equation}
{\cal M}_R = \Omega^{\ast} \text{diag}(M_1,M_2,M_3) \Omega^{\dagger},\label{eq::RHNmassmatrix}
\end{equation}
where $\Omega$ is a unitary matrix and $M_{1,\,2,\,3}\geq 0$. Once the mass matrix of the right-handed neutrinos $\mathcal{M}_R$ is diagonalised, the terms in Eq.~\eqref{eq:LagMinMod} lead to 
\begin{align}
{\Delta \cal L} = 
&-\hat{\lambda}_{j\alpha} \hat{N}_{j}^{c} (\ell_{\alpha} \cdot H) -\frac{1}{2} M_{j} \hat{N}_{j}^{c} \hat{N}_{j}^{c} + \text{h.c.}~,
\label{eq:Lagrangian_MR_diag}
\end{align}
where the sum over equal indices is implicit and 
\begin{eqnarray}
\hat{N}_{j}^{c} &=& \sum_{\alpha} \Omega_{\alpha j}^{\ast} N_{\alpha}^{c}~, \\
\hat{\lambda}_{j\alpha} &=& \Omega_{\alpha j} \lambda_{\alpha}~(\text{not~summed}).
\end{eqnarray}
The Weyl spinor $\hat{N}_{j}$, $j=1,\,2,\,3$, with $(\hat{N}_j^c)_a = \varepsilon_{ab}(\hat{N}_j^*)_b$, describes a right-handed neutrino $N_{j}$ with mass $M_{j}$.

Upon examining Eq.~\eqref{eq:mr} and using the Yukawa coupling parameterisation provided in Eq.~\eqref{eq:lthetaphi}, it becomes transparent that the masses of the right-handed neutrinos are proportional to $\lambda^2$. This dependence  can be expressed more precisely as follows,
\begin{equation}
    M_{1,\,2,\,3} = \frac{v^2\lambda^2}{2 m_1}\, \beta_{1,\,2,\,3}(\theta,\phi)\simeq 6\times 10^{14} \,\text{GeV} \left(\frac{0.05 \,\text{eV}}{m_1}\right)\lambda^2\beta_{1,\,2,\,3}(\theta,\phi),
    \label{eq:mass_scale}
\end{equation}
where $\beta_{1,\,2,\,3}(\theta,\phi)$, satisfying $\beta_{1,\,2,\,3}(\theta,\phi)\lesssim \mathcal{O}(1)$, are some real numbers that only depend on the ratios $m_1/m_2<1$, $m_1/m_3<1$, the PMNS parameters and on trigonometric functions of $\theta$, $\phi$.\footnote{More precisely, $\beta_{1,\,2,\,3}(\theta,\phi)$ are defined by 
$\text{diag}(\beta_1, \beta_2, \beta_3)=-\Omega^T D_{\theta,\phi}UD_mU^T D_{\theta,\phi} \Omega$, where $D_{\theta,\phi}\equiv \text{diag}(\cos\theta,\sin\theta\cos\phi,\sin\theta\sin\phi)$ and $D_m\equiv\text{diag}(1,m_1/m_2,m_1/m_3)$.}
%
\begin{figure}[t!]
    \centering
    \includegraphics[width = .52\textwidth]{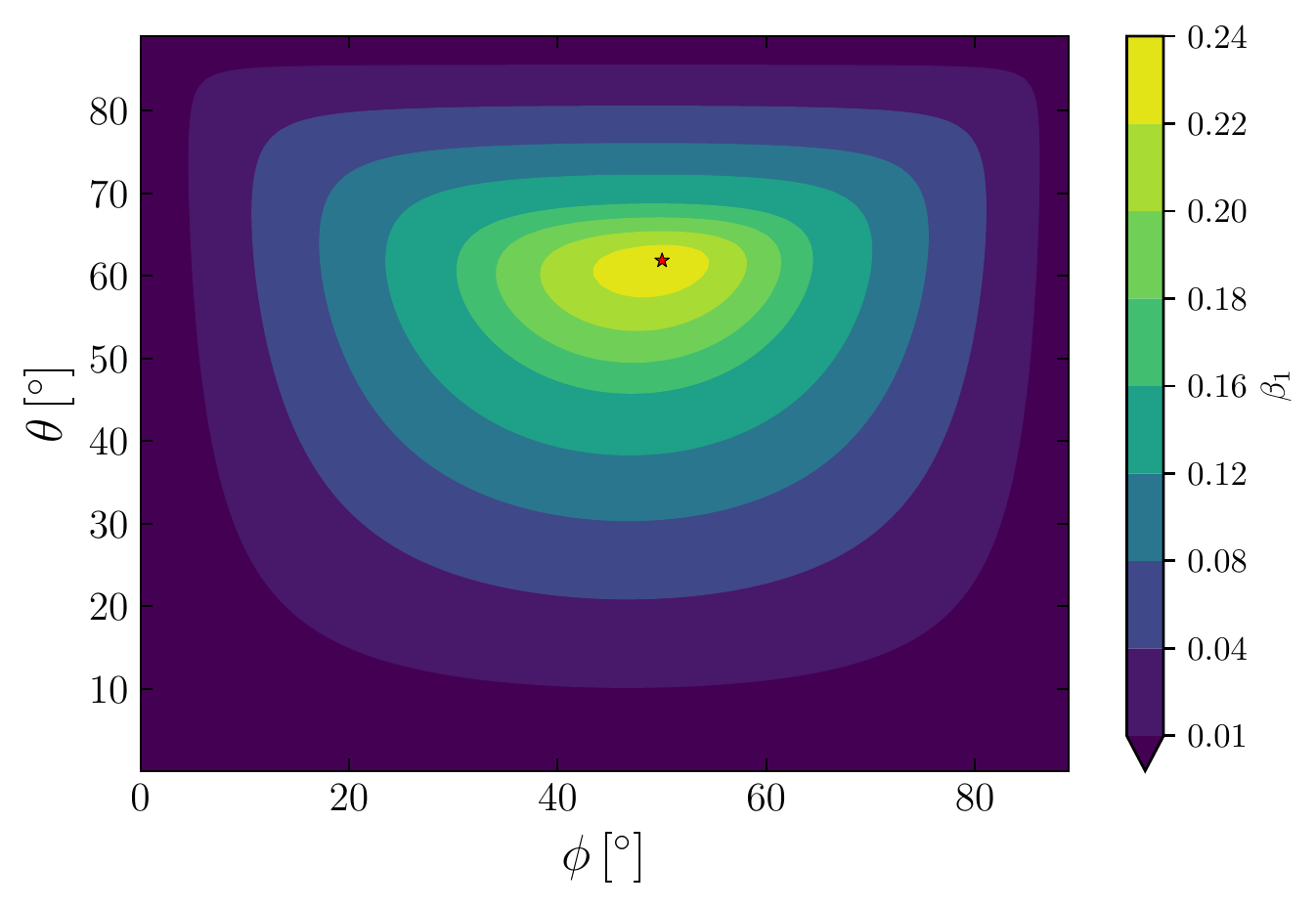}
    \includegraphics[width = \textwidth]{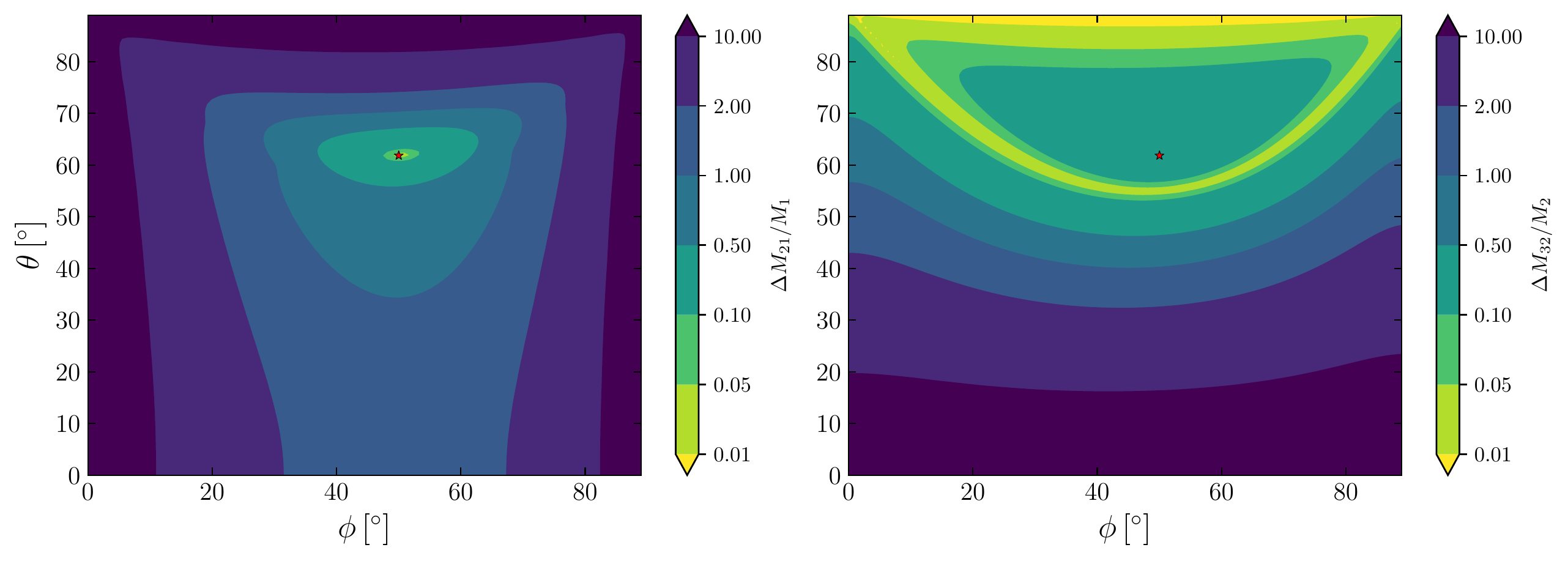}
    \caption{The contour plots for the quantities $\beta_1$ (top panel), $\Delta M_{21}/M_1$ (bottom-left panel) and $\Delta M_{32}/M_2$ (bottom-right panel) in the $\phi-\theta$ plane. The red stars in the plots mark the point of coordinates $\phi\sim 50^\circ$ and $\theta \sim 62^\circ$ for which 
    $\beta_1$
    is maximised  ($\Delta M_{21}/M_1$ is minimised). 
    The plot is obtained for the neutrino mixing angles, and squared mass differences as in Set I. Choosing the input parameters as in Set II would lead to a qualitatively similar figure.}
    \label{fig:Contour_Mass_ratios_withSK}
\end{figure}

Fig.~\ref{fig:Contour_Mass_ratios_withSK} shows the contours for three parameters: $\beta_1=2m_1v^{-2}\lambda^{-2}M_1$ (top), $\beta_2/\beta_1 - 1 = \Delta M_{21}/M_1$ (bottom-left), and $\beta_3/\beta_2-1 = \Delta M_{32}/M_2$ (bottom-right); here, $\Delta M_{21(32)} \equiv M_{2(3)} - M_{1(2)}$. These contours are  obtained using the neutrino mixing angles and squared mass differences as in Set I, with $\phi$ and $\theta$ varied in the $\pi/2$ range with $0 \leq \theta,\, \phi \leq \pi/2$.
We observe that the parameter $\beta_1$ exhibits a maximum value of $\beta_1^\text{max} \sim 0.24$ at the location with coordinates $\phi = \tilde{\phi} \sim 50^\circ$ and $\theta = \tilde{\theta} \sim 62^\circ$, which is marked with a red star. At the same coordinates, we find numerically that $\Delta M_{21}/M_1$ is minimised, while $\Delta M_{32}/M_2$ is locally maximised. 
Moreover, we find that the overall trend of the parameter space remains consistent if we choose to work with the values for $\theta_{12}$, $\theta_{13}$, $\Delta m^2_{21}$ and $\Delta m_{31}^2$ as in Set II. However, the maximum value of $\beta_1$ is found for different values of $\phi$ and $\theta$. Specifically, the maximum value of $\beta_1$ is found at $\phi=\tilde{\phi} \sim 38^\circ$ and $\theta=\tilde{\theta} \sim 65^\circ$, taking a value of $\sim 0.21$. 
Upon examining the bottom panels of Fig.~\ref{fig:Contour_Mass_ratios_withSK}, it is evident that our model predicts a nearly degenerate spectrum for the right-handed neutrinos, with $M_2\lesssim 3M_1$ and $M_3\lesssim 3M_2$, over a large part of the parameter space.  However, we do not observe all three right-handed neutrinos to be simultaneously nearly-degenerate in mass in any region of the parameter space. 
There exists a tiny region where $N_1$ and $N_2$ are nearly-degenerate with  $M_2\lesssim 1.05 M_1$. For $N_2$ and $N_3$, there exists a larger region  nearly degenerate, with $M_3 \lesssim 1.05 M_2$. These regions are shown in yellow-ish colour in the bottom panels of Fig.~\ref{fig:Contour_Mass_ratios_withSK}.\footnote{We suspect that a specific point for which $M_1 = M_2$ exists around $(\tilde{\phi},\tilde{\theta})$ (although finding such precise point would require infinite numerical resolution). It should be possible to consider $M_1\simeq M_2$ with arbitrary precision around that point, but this would require a certain amount of fine-tuning in the choice of the parameters $\theta$ and $\phi$, much more severe than, e.g., the uncertainties in the neutrino parameters we have considered.}

\section{Thermal leptogenesis}
\label{sec:LG}

In this section, we present the equations necessary to investigate the thermal leptogenesis mechanism in the minimal gauged U(1)$_{L_\mu-L_\tau}$ model. To simplify our analysis, we make the following assumptions: i) the U(1)$_{L_\mu-L_\tau}$ gauge symmetry is never restored after the reheating; ii) the masses of the U(1)$_{L_\mu-L_\tau}$ gauge boson and the singlet scalar field associated with $\sigma$ are larger than the reheating temperature $T_R$ so that these fields are always absent from the thermal bath; iii) the masses of all three right-handed neutrinos are smaller than the reheating temperature. The first two can be realised by taking the U(1)$_{L_\mu-L_\tau}$ gauge coupling and the self coupling of the $\sigma$ field sufficiently large and  $\langle \sigma \rangle \gg T_R$. The third assumption indicates $|M_{ee, \mu \tau}|, |\lambda_{e\mu, e\tau} \langle \sigma \rangle| < T_R$.

To take the impact of charged lepton and heavy neutrino flavours into account, we employ the formalism of DMEs instead of the simpler BEs.\footnote{Charged lepton and heavy neutrino flavour effects can be particularly relevant when the right-handed neutrino mass is not strongly hierarchical \cite{Blanchet_2013}, as in our model (see Fig.~\ref{fig:Contour_Mass_ratios_withSK}). In addition, as we will show later, to explain the observed BAU, the mass of the lightest right-handed neutrino can be as low as $M_1\sim 10^{11-12}\,\text{GeV}$, with leptogenesis occurring at the transition between the single- and two-flavour regimes, where BEs fail \cite{Moffat:2018smo, Granelli:2020ysj}. Therefore, it is necessary to solve the DMEs with three decaying right-handed neutrinos and fully take flavour effects into account.}
Specifically, in the case of the three decaying right-handed neutrinos, we can express the DMEs in the following compact form~\cite{Simone_2007, Blanchet_2007, Blanchet_2013, Moffat:2018wke, Moffat:2018smo, Granelli:2021fyc}:
\begin{eqnarray}
\label{DME:N}
\frac{dN_{N_{j}}}{dz}&=&-D_{j}(N_{N_{j}}-N^\text{eq}_{N_{j}}) \,, \\
\label{DME:full3}
 \frac{dN_{\alpha\beta}}{dz} &=&
 \begin{aligned}[t]
 &\sum_{j=1}^3\left[\epsilon^{(j)}_{\alpha\beta}D_{j}(N^{}_{N_{j}}-N^\text{eq}_{N_{j}}) - 
\frac{1}{2}W_{j}\left\{P^{0(j)},N\right\}_{\alpha\beta}\right] \\
-&\frac{\Gamma_\tau}{Hz}
\left[I_\tau,\left[I_\tau,N\right]\right]_{\alpha\beta}
-\,\frac{\Gamma_\mu}{Hz}\left[I_\mu,\left[I_\mu,N\right]\right]_{\alpha\beta}\,,
\end{aligned}
\end{eqnarray}
where the indices $j=1,\,2,\,3$ refer to the three decaying right-handed neutrinos, while $\alpha,\,\beta=e,\,\mu,\,\tau$ to the charged lepton flavours. The variable $z$ is defined as  $z\equiv M_1 / T$.
The quantity $N_{N_j}$ represents the number of heavy neutrinos $N_j$ in a comoving volume, which we normalise using the same method as in Refs.~\cite{Moffat:2018wke,Moffat:2018smo,Granelli:2021fyc}, such that it contains a single photon when $z\ll 1$, i.e., $N^\text{eq}_{N_j}(0) = 3/4$. 
We approximate the equilibrium number density of heavy neutrinos as $N^\text{eq}_{N_j}(z) = (3/8) x_j z^2 K_2(\sqrt{x_j} z)$, where $x_j\equiv (M_j/M_1)^2$ and $K_n(z)$, $n = 1\,,2\,,\,...$, is the modified $n^\text{th}$ Bessel function of the second kind~\cite{BUCHMULLER2005305, Hahn-Woernle:2009jyb}. 
The diagonal entries $N_{\alpha\alpha}$ of the density matrix $N$ correspond to the comoving number densities 
for the $(1/3)B-L_\alpha$ asymmetry, such  that $N_{B-L} = \sum_{\alpha=e,\,\mu,\,\tau} N_{\alpha\alpha}$.
The off-diagonal elements $N_{\alpha\beta}$, where $\alpha\neq \beta$, represent the degree of coherence between the flavour states. 

The projection matrices in the anti-commutator term in the first line of Eq.~\eqref{DME:full3} is defined as,
\begin{equation}
P^{0(j)}_{\alpha \beta} \equiv \frac{\hat{\lambda}_{j \alpha}^* \hat{\lambda}_{j \beta}}{(\hat{\lambda}\hat{\lambda}^\dagger)_{jj}}\,,
\end{equation}
generalising the notion of the projection probability. Furthermore, the double commutator structure in the second line of Eq.~\eqref{DME:full3} involves  $3\times3$ matrices $I_{\tau}$ and $I_\mu$ defined such that $(I_\tau)_{\alpha\beta} = \delta_{\alpha\tau}\delta_{\beta\tau}$ and $(I_{\mu})_{\alpha\beta} = \delta_{\alpha \mu}\delta_{\beta \mu}$.
Additionally, we use the following analytical expressions for $D_j$ and $W_j$~\cite{BUCHMULLER2005305, Moffat:2018wke},
\begin{eqnarray}
\label{eq:Dj}
D_j(z) &=& \kappa_j x_j z \frac{K_1(\sqrt{x_j} z)}{K_2(\sqrt{x_j} z)}\,,\\
W_j(z) &=& \frac{1}{4}\kappa_jx_j^2z^3K_1(\sqrt{x_j}z)\,,
\label{eq:Wj}
\end{eqnarray}
%
where the decay parameter $\kappa_j$ quantifies the strength of the wash-out processes in erasing the asymmetry and is defined as the ratio between the total decay rate of $N_j$ at zero temperature, $\Gamma_{j} = (\hat{\lambda}\hat{\lambda}^\dagger)_{jj} M_j/8\pi$, and the Hubble expansion rate $H$ at $T = M_j$. Note that the decay rate $\Gamma_j$ and the Hubble expansion rate $H(T=M_j)\propto M_j^2$ scale as $\lambda^4$. As a result, the decay parameter $\kappa_j$ depends only on the angles $\theta$ and $\phi$. 
Numerical calculations show that, for any $0\leq\theta,\,\phi\leq \pi/2$ and $j=1,\,2,\,3$, $\kappa_j\gg 1$ and leptogenesis occurs in the strong wash-out regime \footnote{ More specifically, we find numerically that $\kappa_i \gtrsim 40$ and can reach maximal values of $\kappa_{1} \simeq 1.66\times10^2$ and $ \kappa_{2,3}\simeq 1.41\times 10^4$.}. 

To solve the DMEs, we need to specify the right-handed neutrino abundances and lepton asymmetries at the starting point of leptogenesis $z_\text{in}$. The strong wash-out condition $\kappa_i\gg 1$ itself does not guarantee the independence of the initial conditions. With strong wash-outs occurring for each lepton flavour, the final BAU is not affected by different choices of the initial abundances \cite{Blanchet:2006be}. Nonetheless,  it can still be, for instance, that one or two lepton flavours are weakly coupled, and the independence of the initial condition is not guaranteed \cite{Garbrecht:2014bfa}. While it is natural to assume that the Universe was symmetric in the lepton and baryon numbers at the beginning of leptogenesis, with all the entries of the flavour density matrix equal to zero, there is more freedom in the choice of the initial right-handed neutrino abundances. In the main analysis, we focus on the case for which the starting abundances of the right-handed neutrinos are zero, as this situation is generically more natural in cosmological inflation models. 
We dedicate Appendix \ref{App:TIA} to discuss how our results change with thermal initial abundances.

It has also been estimated that, in the strong wash-out regime and in the case of a hierarchical right-handed neutrino mass spectrum, the $\Delta L = 1$ scattering processes and related wash-outs would only contribute to the final BAU at most by $\mathcal{O}(10\%)$ \cite{Blanchet:2006be, Frossard:2013bra}. We find that this is still true in most of the parameter space of our model, except for certain fine-tuned choices of the parameters. Therefore, we do not consider the effects of scatterings in the main analysis \footnote{We are neglecting from our analysis the contributions to the BAU from the early period when the right-handed neutrinos are relativistic and the effects of spectator processes that could potentially affect our final results (see, e.g., \cite{Garbrecht:2019zaa} and references therein). We leave the implementation of these effects to future related studies.}.

The CP-asymmetry $\epsilon_{\alpha\beta}^{(j)}$ appearing in the DMEs can be separated into contributions from vertex and self-energy diagrams, namely $\epsilon_{\alpha\beta}^{(j)}\equiv \epsilon_{\alpha\beta}^{\text{V}\,(j)} + \epsilon_{\alpha\beta}^{\text{S}\,(j)}$. The two contributions are given by
\cite{Flanz:1994yx,Covi:1996wh,Covi:1996fm,Buchmuller:1997yu,Abada:2006fw,Simone_2007,Blanchet_2013,Biondini_2018},
\begin{eqnarray}
\label{eq:epsV}
\epsilon^{\text{V}\,(j)}_{\alpha\beta}&=&\frac{1}{16\pi\left(\hat{\lambda}\hat{\lambda}^{\dagger}\right)_{jj}}
\sum_{k\neq j}\left\{ i\left[\hat{\lambda}_{j\alpha}^*\hat{\lambda}_{k\beta}(\hat{\lambda}\hat{\lambda}^{\dagger})_{kj}
- \hat{\lambda}_{j\beta}\hat{\lambda}_{k\alpha}^*(\hat{\lambda}\hat{\lambda}^{\dagger})_{jk}\right] \xi\left(x_k/x_j\right)\right\} \,,\\
\label{eq:epsS}
\epsilon^{\text{S}\,(j)}_{\alpha\beta}&=&\frac{1}{16\pi\left(\hat{\lambda}\hat{\lambda}^{\dagger}\right)_{jj}}
\sum_{k\neq j}\Bigg\{i\left[\hat{\lambda}_{j\alpha}^*\hat{\lambda}_{k\beta}(\hat{\lambda}\hat{\lambda}^{\dagger})_{kj}
- \hat{\lambda}_{j\beta}\hat{\lambda}_{k\alpha}^*(\hat{\lambda}\hat{\lambda}^{\dagger})_{jk}\right] \frac{\sqrt{x_k/x_j}}{x_k/x_j-1}  \\\nonumber
&&~~~~~~~~~~~~~~~~~~~~~~+
i\left[\hat{\lambda}_{j\alpha}^*\hat{\lambda}_{k\beta}(\hat{\lambda}\hat{\lambda}^{\dagger})_{jk}-\hat{\lambda}_{j\beta}\hat{\lambda}^*_{k\alpha}(\hat{\lambda}\hat{\lambda}^{\dagger})_{kj}\right] \frac{1}{x_k/x_j-1} \Bigg\}\,,
\end{eqnarray}
where $\xi(x)\equiv \sqrt{x} \left[(1+x)\log\left(1+1/x\right)-1\right]$. 
In the case of a degeneracy in the right-handed neutrino mass spectrum, the self-energy contribution to the CP-asymmetry is resonantly enhanced and can dominate over the vertex contribution. However, the self-energy contribution in Eq.~\eqref{eq:epsS} becomes ill-defined when two neutrinos, say $N_j$ and $N_k$, are quasi-degenerate in mass, as $1/(x_k/x_j-1)\to \infty$ when $x_k/x_j\to 1$. 
To address this non-physical behaviour, the full-resummed Yukawa couplings must be considered in the calculations, and the self-energy contribution to the CP-asymmetry can be regularised by performing the following substitution \cite{Pilaftsis:1997jf, Pilaftsis:1997dr, Pilaftsis:2003gt, Pilaftsis:2005rv}, \begin{equation}\label{eq:resonance_reg}
\frac{1}{x_k/x_j - 1} \to \frac{(M_k^2-M_j^2)M_j^2}{(M_k^2-M_j^2)^2 + M_j^4\Gamma_k^2/M_k^2}
\,.
\end{equation}

The regularised CP-asymmetry, obtained by using the prescription in Eq.\eqref{eq:resonance_reg}, does not suffer from divergences and vanishes for equal right-handed neutrino masses. However, the self-energy contribution to the CP-asymmetry, even after regularisation, still exhibits a resonance. In particular, the self-energy contribution is maximised when $|\Delta M_{jk}|\simeq 0.5\,\Gamma_j$, with $\Delta M_{jk}\equiv M_j-M_k$. This resonant behaviour has been extensively studied in the context of resonant leptogenesis, especially in the efforts to avoid the Davidson-Ibarra bound \cite{Davidson:2002qv} and extend the scenario of thermal leptogenesis down to the electroweak scale \cite{Pilaftsis:1997jf,Pilaftsis:1997dr, Hambye:2001eu, Pilaftsis:2003gt, Hambye:2004jf, Pilaftsis:2005rv, Cirigliano:2006nu, Xing:2006ms, Branco:2006hz, Chun:2007vh, Kitabayashi:2007bs, Deppisch:2015qwa, Brivio:2019hrj, Granelli:2020ysj}.  A comprehensive review can be found in Ref.~\cite{Dev:2017wwc}. The ratios $|\Delta M_{jk}|/ \Gamma_j$, $j\neq k$, quantify the importance of resonance effects. Far away from the resonance, when $|M_j-M_k|\gg \Gamma_j$, the effects of the enhancement are sub-leading, and the regularised CP-asymmetry obtained  with the prescription in Eq.~\eqref{eq:resonance_reg} resembles the form given in Eq.~\eqref{eq:epsS}. We perform a scan of $|\Delta M_{12}|/ \Gamma_1$ and $|\Delta M_{23}|/ \Gamma_2$ over the $\phi-\theta$ plane for various choices of $\lambda$. We found  that resonance effects are significant only in certain small regions of the parameter space.  Furthermore, these regions become even smaller with  decreasing values of  $\lambda$ and are effectively negligible when $\lambda \lesssim 0.5$. The  details of this analysis can be found in Appendix \ref{App:resonance}. Nevertheless, we have included the effects of resonance as in Eq.~\eqref{eq:resonance_reg} in our calculations.\footnote{Different approaches to regularise the self-energy contribution to the CP-asymmetry have been proposed~\cite{Garny:2011hg, BhupalDev:2015oxc, Dev:2017wwc}. 
However, in our case,  the mass splittings between the right-handed neutrinos are much larger than their decay widths and we expect that a different estimation of the resonant effects would not significantly affect our results.  It is worth noting that in Ref.~\cite{Pilaftsis:2003gt} (and in the recent study in Ref.~\cite{daSilva:2023zxk}), a more sophisticated regularisation taking into account the mass degeneracy between all the three right-handed neutrinos was found. 
 However, our analysis presented in Appendix \ref{App:resonance} shows no region of the parameter space of our model for which $|M_1-M_2|/\Gamma_1\lesssim 10$ and $|M_2-M_3|/\Gamma_2\lesssim 10$ simultaneously. Hence, for our purposes, it is sufficient to consider the mass degeneracy between two right-handed neutrinos at a time and use the simplified regularisation in Eq.~\eqref{eq:resonance_reg}.}

Finally, we numerically solve the DMEs in Eqs.~\eqref{DME:N} and \eqref{DME:full3} with the Python package \texttt{ULYSSES} \cite{Granelli:2020pim, Granelli:2023vcm}. The code computes $N_{B-L} =  N_{ee} + N_{\mu\mu} +  N_{\tau\tau}$ and relates it to the present baryon-to-photon ratio via
\begin{equation}
    \eta_B = \frac{c_s}{f}
N_{B-L}\approx 0.013 N_{B-L}\,,
\label{eq:etaBl}
\end{equation}
%
where $c_s$ is the SM sphaleron conversion coefficient and 
the $f$ factor comes from the dilution of the baryon asymmetry 
 due to the  change in the photon density  
between leptogenesis and recombination 
\cite{BUCHMULLER2005305}.

\section{Results of the Parameter Scan of Viable Leptogenesis}
\label{sec:results}
In this section, we discuss the results of our parameter scan aimed at identifying the viable space for thermal leptogenesis by solving the set of DMEs 
introduced in Sec.~\ref{sec:LG} in the case of vanishing right-handed neutrino initial abundances.
\begin{figure}[t!]
    \centering
    \includegraphics[width = .48\textwidth]{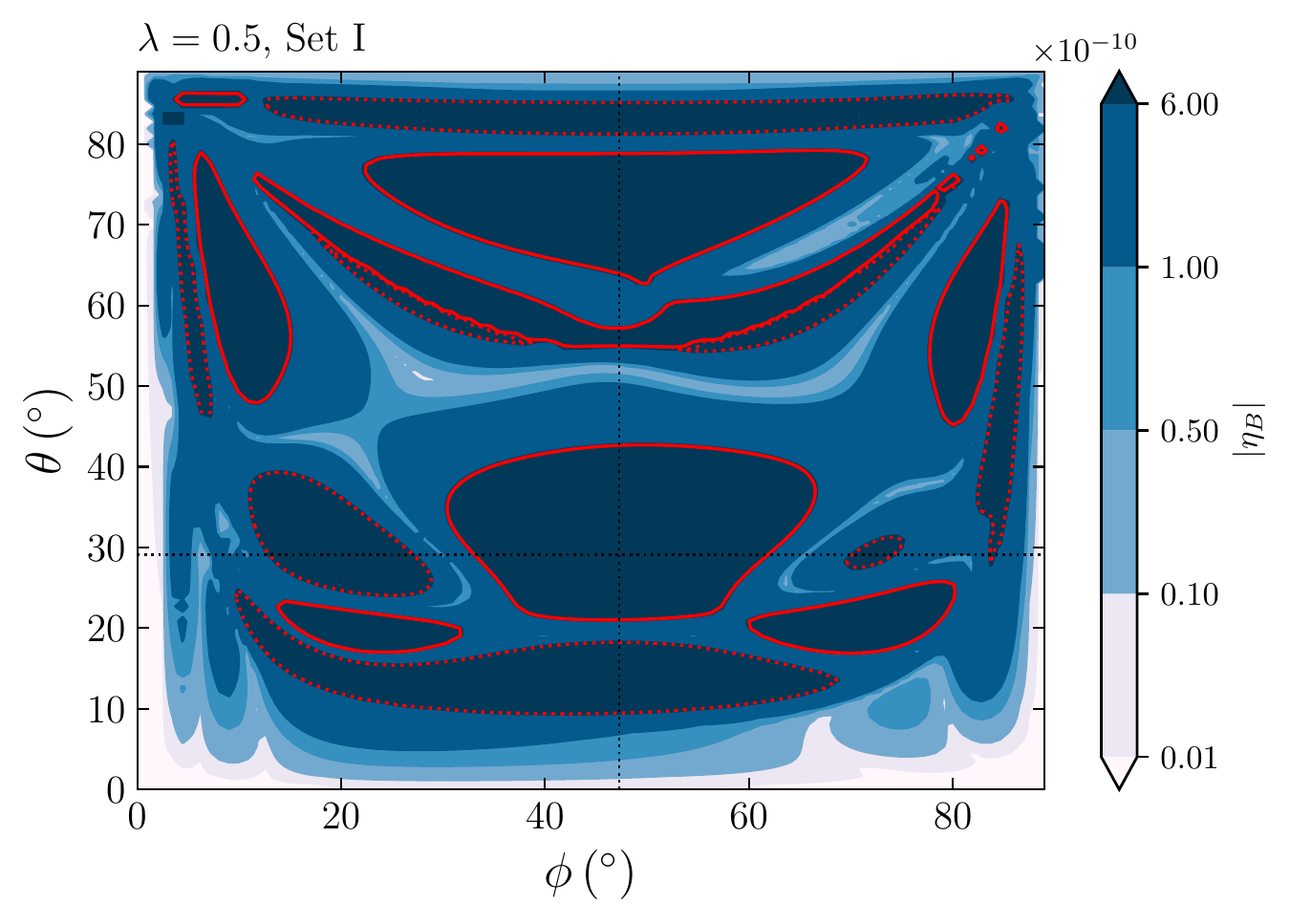}
    \includegraphics[width = .48\textwidth]{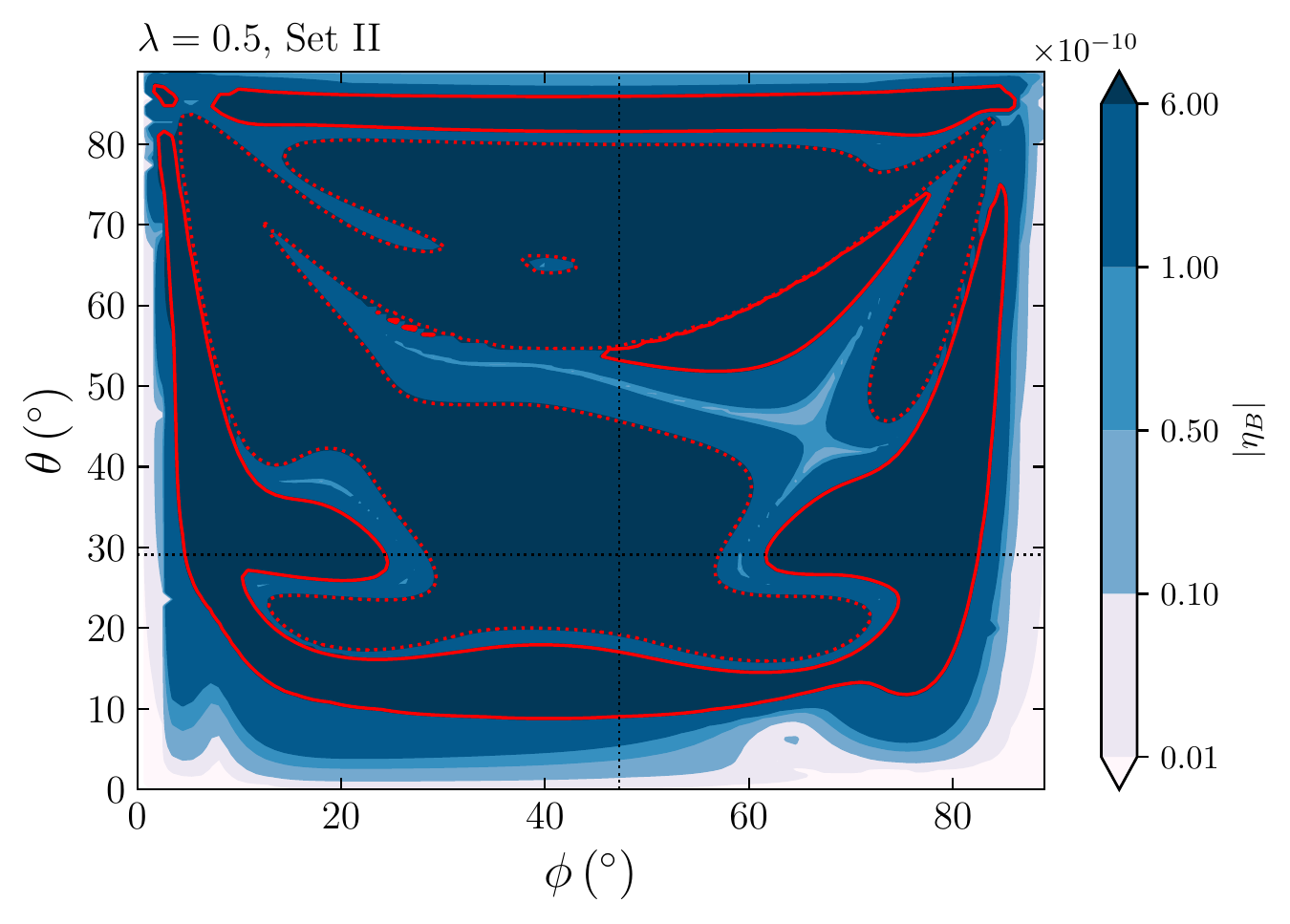}
        \includegraphics[width = .48\textwidth]{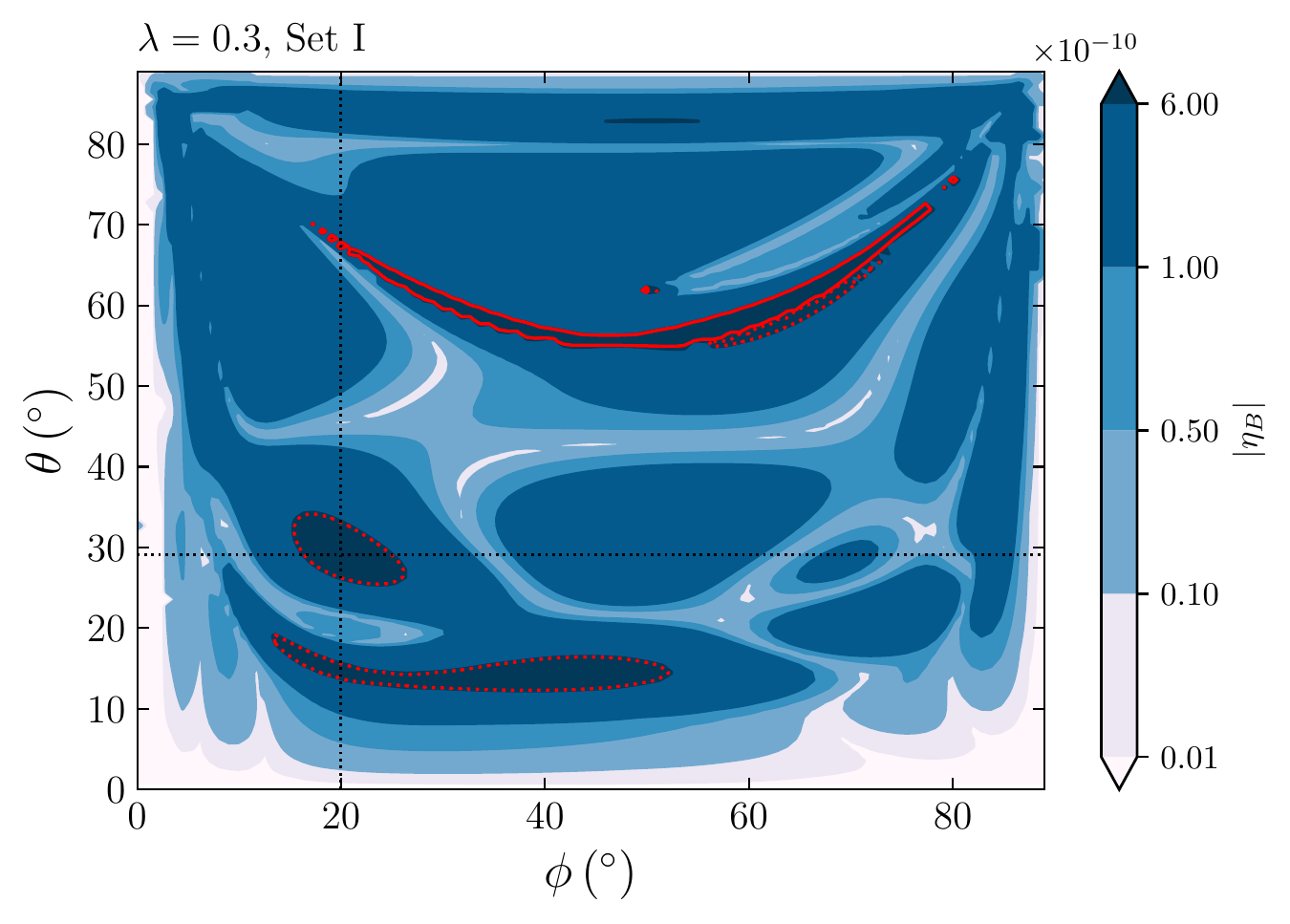}
    \includegraphics[width = .48\textwidth]{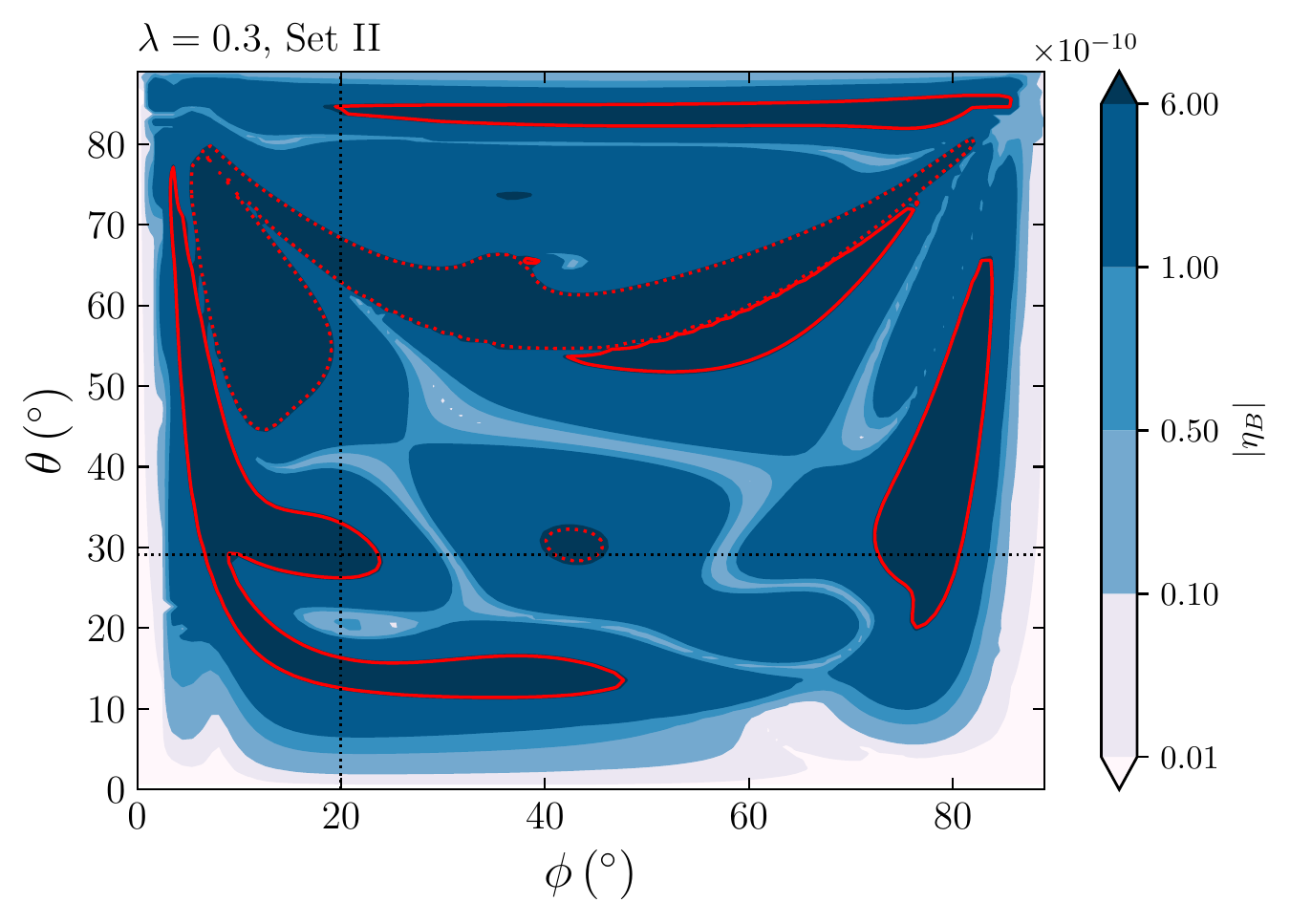}
        \includegraphics[width = .48\textwidth]{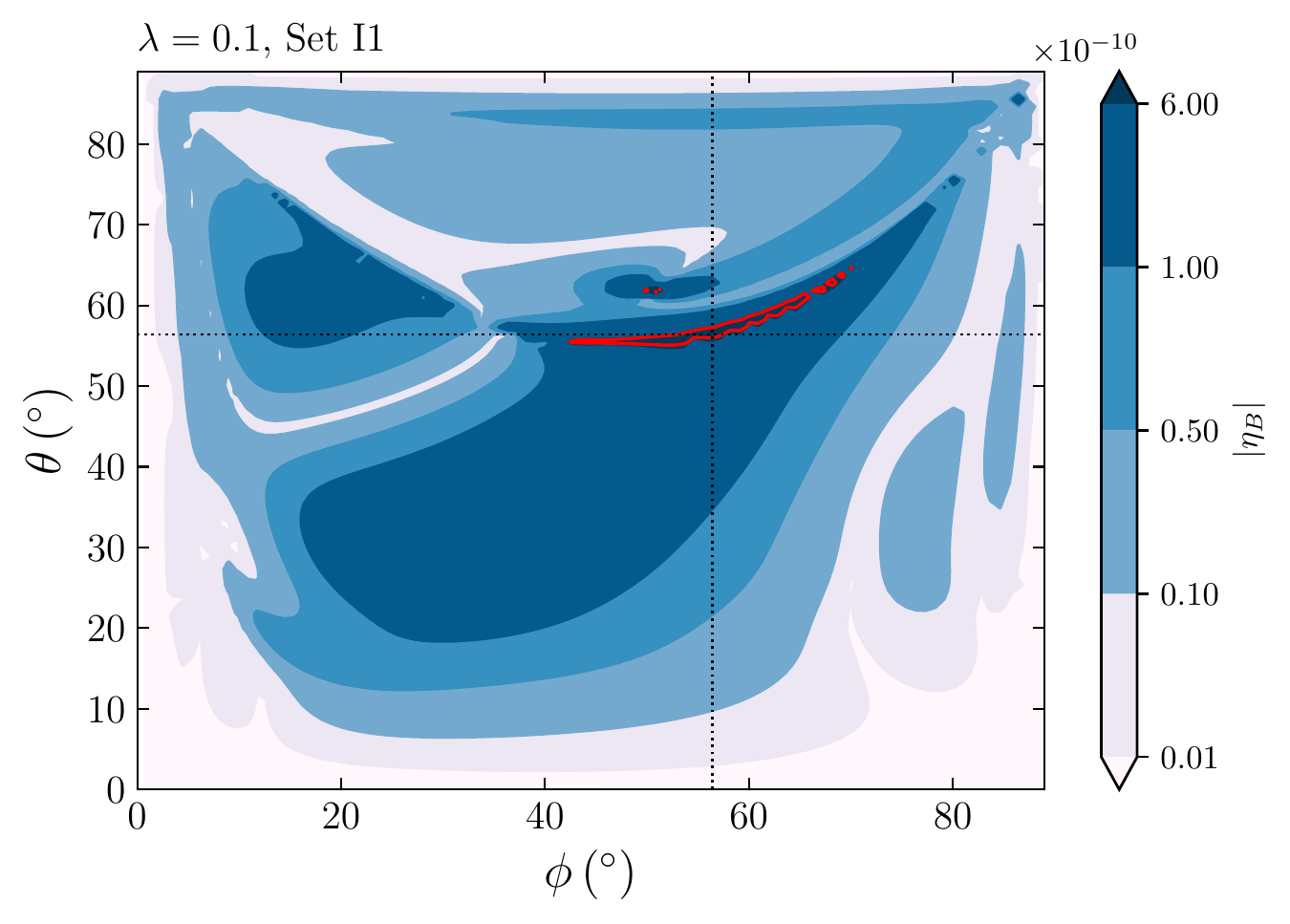}
    \includegraphics[width = .48\textwidth]{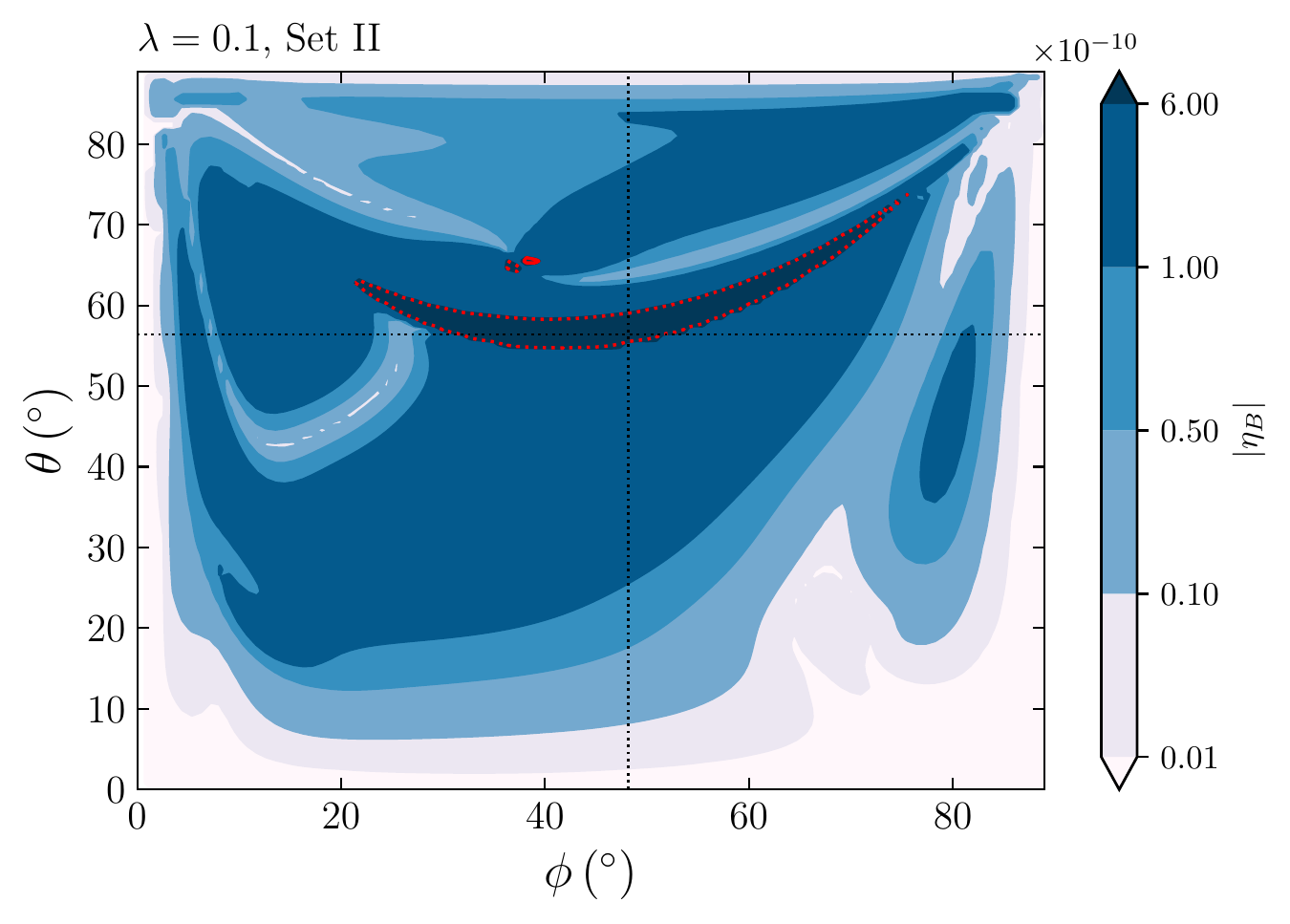}
    \caption{The contour plots in the $\phi-\theta$ plane of the BAU (in modulus) predicted by the DMEs with three decaying right-handed neutrinos in the case of vanishing initial abundances. The plots are obtained for $\lambda = 0.5$ (top panels), $0.3$ (central panels) and $0.1$ (bottom panels), with the input parameters $\theta_{12}$, $\theta_{13}$, $\theta_{23}$, $\Delta m^2_{21}$ and $\Delta m^2_{31}$ as in Set I with $(\delta,\,\alpha_2,\,\alpha_3) = (301^\circ,\,116^\circ,\, 269^\circ)$ (left panels) and Set II with $(\delta,\,\alpha_2,\,\alpha_3) = (228^\circ,\,225^\circ,\, 70^\circ)$ (right panels). As indicated in the bar legends, darker (lighter) regions correspond to larger (smaller) values of the BAU, with the red solid contours representing the points for which the predicted BAU equals the observed value $\eta_B \simeq 6.1\times 10^{-10}$. For the red dotted contours, the BAU equals the observed value in modulus, but the sign is negative. The dotted vertical and horizontal lines mark the benchmark points BMPa (top panels), BMPb (central panels), BMPcI and BMPcII (bottom panels).}
    \label{fig:LG_Scan}
\end{figure}
Fig.~\ref{fig:LG_Scan} shows the results of the scan in the $\phi-\theta$ plane for three benchmark values of $\lambda$, namely $\lambda = 0.5$ (top panels), $\lambda = 0.3$ (central panels) and $\lambda = 0.1$ (bottom panels). The plots in the left  and right panels are obtained with the neutrino mixing angles and squared mass differences as in Set I with $(\delta,\,\alpha_2,\,\alpha_3) = (301^\circ,\,116^\circ,\, 269^\circ)$ and Set II with $(\delta,\,\alpha_2,\,\alpha_3) = (228^\circ,\,225^\circ,\, 70^\circ)$, respectively. The regions of viable leptogenesis are those surrounded by the solid red contours, where the predicted BAU matches the observed value $\eta_B = 6.1\times 10^{-10}$. The points corresponding to the dotted red contours result in a predicted BAU of $-6.1\times 10^{-10}$, indicating the correct magnitude but wrong sign. To darker (lighter) regions correspond larger (smaller) values of the predicted BAU, in modulus. 

\begin{figure}
    \centering
    \includegraphics[width = .48\textwidth]{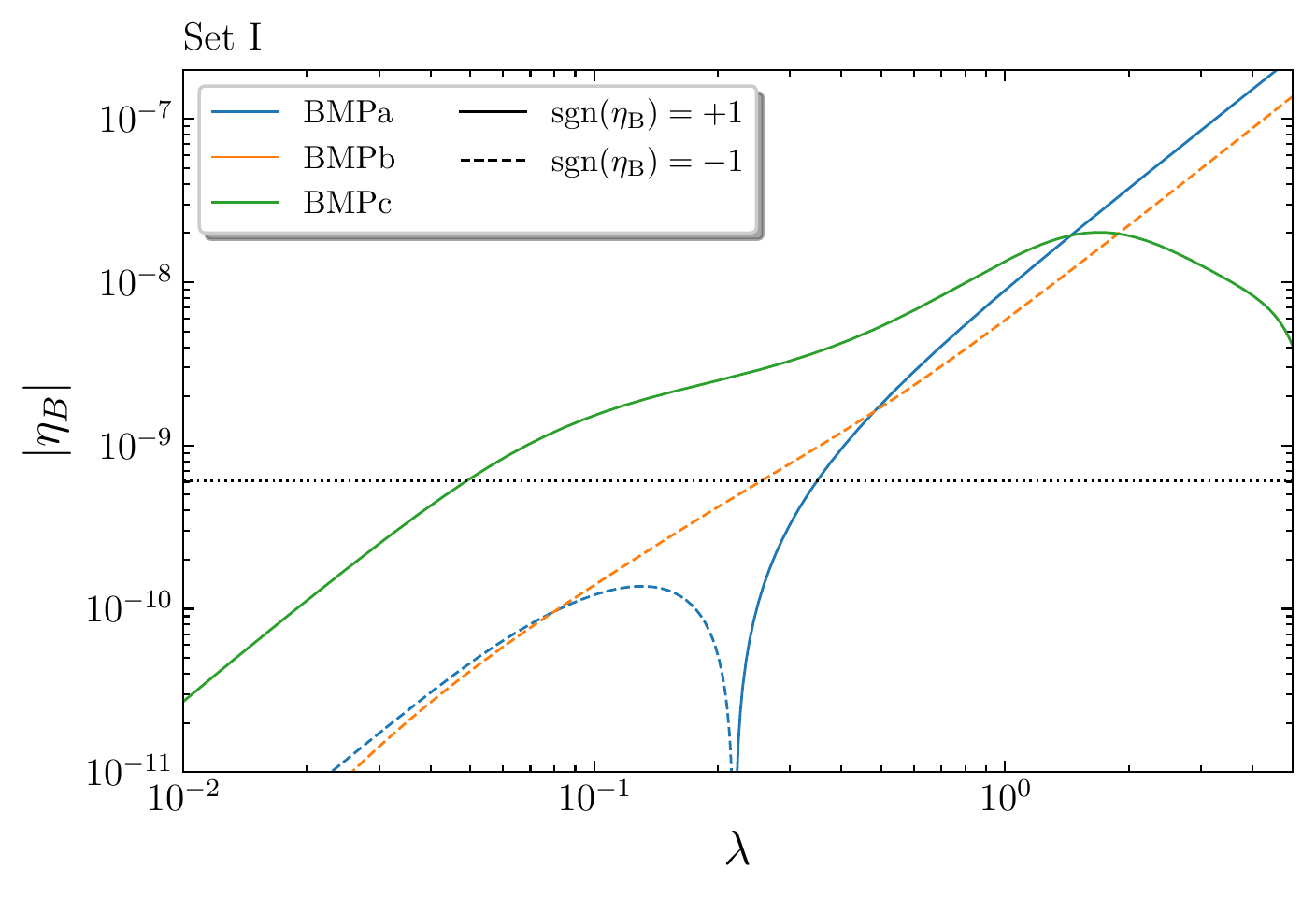}
    \includegraphics[width = .48\textwidth]{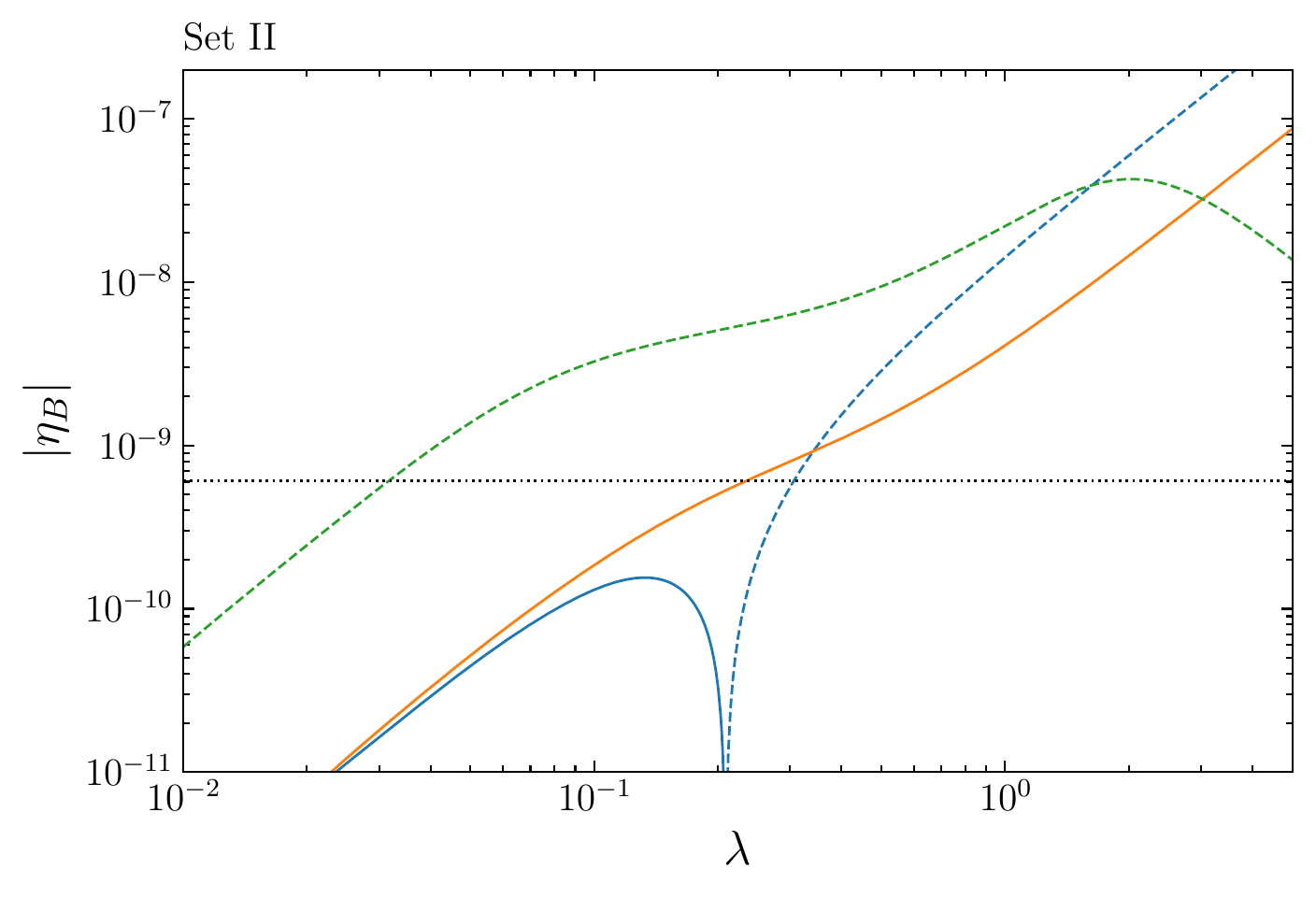}
    \caption{The BAU as a function of $\lambda$ predicted by the DMEs with three decaying right-handed neutrinos. The input parameters in the left and right panels are as in Set I with $(\delta,\,\alpha_2,\,\alpha_3) = (301^\circ,\,116^\circ,\, 269^\circ)$ and Set II with $(\delta,\,\alpha_2,\,\alpha_3) = (228^\circ,\,225^\circ,\, 70^\circ)$, respectively. The different curves are obtained for BMPa (blue), BMPb (orange), BMPcI and BMPcII (green), with the solid (dashed) style indicating the positive (negative) sign of the predicted baryon-to-photon ratio. Note the overall sign flip between the curves obtained in the two different sets of input parameters.}
    \label{fig:BAUvsLam}
\end{figure}

In Fig.~\ref{fig:LG_Scan}, it is clear that  there are numerous regions in the parameter space where leptogenesis within the minimal gauged U(1)$_{L_\mu-L_\tau}$ model predicts a BAU equal to or greater than the observed value, in modulus. However, the sign of the BAU alternates between positive and negative values in different regions, and only those regions where the BAU is positive correspond to successful leptogenesis. It is worth noting that the sign of the BAU can always be changed without changing its magnitude by switching to the equivalent solution with $(m_1,\,2\pi-\delta,\, 2\pi-\alpha_2,\,2\pi-\alpha_3)$ \cite{Asai:2017ryy}. This can be shown as follows. First, note that switching from a set of solutions $(m_1,\, \delta,\, \alpha_2,\, \alpha_3)$ to the other one $(m_1,\, 2\pi-\delta,\, 2\pi-\alpha_2,\, 2\pi-\alpha_3)$ yields $U \to U^*$, and thus $\mathcal{M}_{\nu_L} \to \mathcal{M}_{\nu_L}^*$, $\mathcal{M}_R \to \mathcal{M}_R^*$, $\Omega \to \Omega^*$, and $\hat{\lambda} \to \hat{\lambda}^*$. Then, $\epsilon^{(j)}_{\alpha \beta}$ and $P^{0(j)}_{\alpha\beta}$ in Eq.~\eqref{DME:full3} transform as $\epsilon^{(j)}_{\alpha \beta} \to - \epsilon^{(j)}_{\beta \alpha}$ and $P^{0(j)}_{\alpha\beta} \to P^{0(j)}_{\beta \alpha}$, respectively. This indicates that if $N_{N_j}$ and $N_{\alpha \beta}$ are solutions of Eqs.~\eqref{DME:N} and \eqref{DME:full3} for a given set of $(m_1,\, \delta,\, \alpha_2,\, \alpha_3)$, then the solutions for $(m_1,\, 2\pi-\delta,\, 2\pi-\alpha_2,\, 2\pi-\alpha_3)$ are given by $N_{N_j}$ and $- N_{\beta \alpha}$. By noting that $\eta_B \propto (N_{ee} + N_{\mu\mu} +  N_{\tau\tau})$, we conclude that if the baryon-to-photon ratio for $(m_1,\, \delta,\, \alpha_2,\, \alpha_3)$ is given by $\eta_B$, that for $(m_1,\, 2\pi-\delta,\, 2\pi-\alpha_2,\, 2\pi-\alpha_3)$ is given by $- \eta_B$.   
It is also important to  note that there is an overall sign difference between the parameter scans obtained for the two sets of input parameters, Set I and II. This is because $\theta_{23}$ varies from values below $ 45^\circ$ in Set I to values above $45^\circ$ in Set II, resulting in a shift in the PMNS phases and an opposite sign of the BAU.\footnote{The behaviour is not exactly symmetric in $\theta_{23}$, see Fig.~1(b) of Ref.~\cite{Asai:2020qax}.} Consequently, only precise combinations of the phase $\delta$ and mixing angle $\theta_{23}$ can lead to the correct sign of the BAU in a particular point of the $\phi-\theta$ plane. Determining whether $\theta_{23} <$ or $> \pi/4$ and/or $\delta <$ or $> \pi$ in future experiments would rule out certain regions of the parameter space of viable leptogenesis in our model only based on the sign of the predicted baryon-to-photon ratio.

In addition, the size of the allowed regions for successful leptogenesis in the $\phi-\theta$ plane is dependent on the mass scale of leptogenesis, which goes as $\lambda^2$. As $\lambda$ decreases (increases), the allowed ranges of $\phi$ and $\theta$ for successful leptogenesis become smaller (larger). 
To determine the minimal values of $\lambda$ at which the viable regions shrink to points,  we search for the local maxima of the predicted BAU.
We identify four benchmark points (BMPs) in the  $\phi-\theta$ plane, around which the symmetry is locally maximised. The BMPs are located at the coordinates BMP a) $\theta = 29.06^\circ$ and $\phi = 47.28^\circ$; BMP b) $\theta = 29.06^\circ$ and $\phi = 19.94^\circ$; BMP cI) $\theta = 56.39^\circ$ and $\phi = 56.39^\circ$ for Set I, and BMP cII) $\theta = 56.39^\circ$ and $\phi = 48.19^\circ$ for Set II. These four BMPs are indicated in Fig.~\ref{fig:LG_Scan} by horizontal and vertical dotted grid lines.

Finally, Fig.~\ref{fig:BAUvsLam} illustrates  the dependence of the BAU  on the scale $\lambda$ for the four different BMPs. The left and right panels correspond to input parameters as in Set I with $(\delta,\,\alpha_2,\,\alpha_3) = (301^\circ,\,116^\circ,\, 269^\circ)$ and Set II with $(\delta,\,\alpha_2,\,\alpha_3) = (228^\circ,\,225^\circ,\, 70^\circ)$, respectively. 
Our numerical analysis yields the following minimal values of $\lambda$ for which leptogenesis is viable: for Set I,  
$\lambda \simeq 0.35$ ($M_1\simeq 10^{12.8}\,\text{GeV}$) for BMPa 
and $(\delta,\,\alpha_2,\,\alpha_3) = (301^\circ,\,116^\circ,\, 269^\circ)$; $\lambda \simeq 0.25$ ($M_1\!\simeq \!10^{12.2}\,\text{GeV}$) for BMPb and $(\delta,\alpha_2,\alpha_3) \!=\! (59^\circ,\,244^\circ,\, 91^\circ)$; $\lambda \simeq 0.05$ ($M_1\simeq 10^{11.6}\,\text{GeV}$) for BMPcI and $(\delta,\,\alpha_2,\,\alpha_3) = (301^\circ,\,116^\circ,\, 269^\circ)$. For Set II, we find: $\lambda \simeq 0.305$ ($M_1\simeq 10^{12.8}\,\text{GeV}$) for BMPa and $(\delta,\,\alpha_2,\,\alpha_3) = (132^\circ,\,135^\circ,\, 290^\circ)$; $\lambda \simeq 0.25$ ($M_1\simeq 10^{12.3}\,\text{GeV}$) for BMPb and $(\delta,\,\alpha_2,\,\alpha_3) = (228^\circ,\,225^\circ,\, 70^\circ)$; $\lambda \simeq 0.03$ ($M_1\simeq 10^{11.2}\,\text{GeV}$) for BMPcII and $(\delta,\,\alpha_2,\,\alpha_3) = (228^\circ,\,225^\circ,\, 70^\circ)$. Overall, we find that leptogenesis is viable for $M_1\gtrsim 10^{11}\,\text{GeV}$ across the entire parameter space. Qualitatively similar results hold in the case of thermal initial abundances of the right-handed neutrinos. See Appendix \ref{App:TIA} for more details.

\section{Conclusions}
\label{sec:conclusions}

In this work, we have studied thermal leptogenesis in the context of the minimal gauged U(1)$_{L_\mu-L_\tau}$ model.
Given the light neutrino squared mass differences and the PMNS mixing angles as input, the model predicts the values of the lightest neutrino mass $m_1$, as well as the PMNS phases $\delta$, $\alpha_2$, and $\alpha_3$. The model remains with three free parameters, which we have identified as $\lambda$, $\theta$, and $\phi$ according to the parameterisation of the Yukawa couplings as in Eq.~\eqref{eq:lthetaphi}.
We have then performed a numerical scan of the parameter space and searched for the allowed ranges of $\lambda$, $\theta$ and $\phi$ for which leptogenesis is viable in reproducing the observed value of the baryon asymmetry of the Universe.
To fully account for the effects of the charged lepton and heavy neutrino flavours, we have solved numerically the sets of density matrix equations instead of the simpler Boltzmann equations, and took the decays of all three right-handed neutrinos into account. Noting that thermal leptogenesis proceeds in the strong wash-out regime in the entire parameter space of the model so that the effects of scatterings and different initial conditions are typically sub-leading, we have focused only on direct and inverse decays and on the case of vanishing initial right-handed neutrino abundances. To avoid the non-physical behaviour in the CP-asymmetry due to degeneracy in the right-handed neutrino mass spectrum, we have included the full-resummed Yukawa couplings and resonant effects in the calculation, even though we found those to be relevant only in secluded regions of the parameter space.

We have found numerous regions in the parameter space where thermal leptogenesis within the minimal gauged U(1)$_{L_\mu-L_\tau}$ model predicts the correct baryon asymmetry of the Universe. This result was not obvious given the restrictions in the neutrino mass structure imposed by the model. The size of the allowed regions for successful leptogenesis in the $\phi-\theta$ plane decreases with the mass scale of leptogenesis, which goes as $\lambda^2$, implying minimal values of the mass scale and $\lambda$ for which leptogenesis can be successful. We found that thermal leptogenesis is viable for $M_1\gtrsim 10^{11-12}\,\text{GeV}$ across the entire parameter space, with $\lambda$ taking values from order unity down to $\mathcal{O}(0.03-0.05)$. These values are larger than those obtained in the context of non-thermal leptogenesis within the minimal gauged U(1)$_{L_\mu-L_\tau}$ model~\cite{Asai:2020qax}. The difference is mostly due to the wash-out effects that are present in the thermal leptogenesis mechanism but not in the non-thermal scenario, implying relatively heavier right-handed neutrinos to satisfy the out-of-equilibrium condition.

The sign of the baryon asymmetry can be either positive or negative in the various regions, but leptogenesis is successful only where the baryon asymmetry is positive. However, after specifying the neutrino squared mass differences and the PMNS mixing angles, the minimal gauged U(1)$_{L_\mu - L_\tau}$ model predicts two distinct sets of PMNS phases, $(\delta,\,\alpha_2,\,\alpha_3)$ and $(2\pi-\delta,\,2\pi-\alpha_2,\,2\pi-\alpha_3)$, each corresponding to opposite signs of the predicted baryon asymmetry. At present, the uncertainty on the estimate of $\delta$ obtained in the global analysis \cite{Esteban:2020cvm} is relatively large so that, in general, both predictions are plausible (to a certain level of confidence). We are then allowed to change the sign of the predicted asymmetry by switching from one set of PMNS phases to the other. The baryon asymmetry also has an opposite sign depending on whether $\theta_{23}$ lies above or below $\pi/4$. According to the $3\sigma$ ranges for $\theta_{23}$ obtained in the global analysis \cite{Esteban:2020cvm}, see Table \ref{Tab::BestFit}, both cases are still valid, and it is, therefore, impossible to discriminate the sign of the baryon asymmetry in a given region of the parameter space. Of course, with future more accurate measurements of the Dirac phase $\delta$ and of the mixing angle $\theta_{23}$ from, e.g., T2K \cite{T2K:2011qtm}, NO$\nu$A \cite{NOvA:2021nfi}, 
DUNE \cite{DUNE:2021tad}, and Hyper-Kamiokande \cite{Hyper-Kamiokande:2022smq} (see also Ref.~\cite{Huber:2022lpm}), we would be able to rule out the certain region of the parameter space of leptogenesis in the considered model only on the basis of the sign of the baryon asymmetry. We stress that the results of this work are obtained for two different extreme choices of $\theta_{23}$ corresponding to the $\pm 3\sigma$ limit values of the global analysis, so to minimize the sum of neutrino masses, $\sum_i m_i$, and the tension with the cosmological bounds. In the future, more precise measurements of $\theta_{23}$ and/or $\sum_i m_i$ will likely impose even more stringent constraints on the model presented here.

\section*{Acknowledgements}
A.G.~wishes to thank the Kavli IPMU and the Department of Physics of the University of Tokyo at Hongo Campus for the kind hospitality offered during the first part of this project. 
A.G.~acknowledges the use of computational resources from the parallel computing cluster of the Open Physics Hub (\small{\href{https://site.unibo.it/openphysicshub/en}{https://site.unibo.it/}\href{https://site.unibo.it/openphysicshub/en}{openphysicshub/en}) at the Physics and Astronomy Department in Bologna. 
The work of A.G.~has received funding from the European Union’s Horizon Europe research and innovation programme under the Marie Skłodowska-Curie Staff Exchange grant agreement No.~101086085 - ASYMMETRY. 
The work of K.H., N.N., and M.R.Q. was supported in part by the Grant-in-Aid for Innovative Areas (No.~19H05810 [K.H.], No.~19H05802 [K.H.], No.~18H05542 [N.N.]), Scientific Research B (No.~20H01897 [K.H., N.N., and M.R.Q.]), and Young Scientists (No.~21K13916 [N.N.]).
The work of J.W.~is supported by the JSPS KAKENHI Grant (No.~22J21260).

\appendix

\section{The Case of Thermal Initial Abundance}\label{App:TIA}

We show in Fig.~\ref{fig:LG_Scan_TIA} the results of the parameter scan in the case of right-handed neutrino thermal initial abundances, i.e.,~$N_{N_j}(z_\text{in}) = N_{N_j}^\text{eq}(0) = 3/4$, $j=1,\,2,\,3$. The plots in the figure reveal some differences with respect to the vanishing initial abundance case in Fig.~\ref{fig:LG_Scan}, indicating a dependence on the initial conditions in the scenario considered in this work. The differences, however, appear for larger values of $\lambda$ and concentrate in the outer part of the $\phi-\theta$ plane. In these regions, given the parametrisation of the Yukawa couplings in Eq.~\eqref{eq:lthetaphi}, we can have that one of the flavour is weakly coupled with either $\lambda_{e,\,\mu,\,\tau}\ll \lambda$, avoiding the wash-outs processes even if the overall strong regime holds, i.e.,~$\kappa_{1,\,2,\,3}\gg 1$. As $\lambda$ takes smaller values, the regions of viable leptogenesis reduce in size and concentrate in the inner regions of the parameter space. For $\lambda = 0.3$, the regions are more similar to those obtained in the vanishing initial abundance case, with no prominent dependence on the initial conditions. This means that, in terms of allowed ranges of masses and couplings, we obtain qualitatively similar results as in the vanishing initial abundance case. 

\begin{figure}
    \centering
    \includegraphics[width = .48\textwidth]{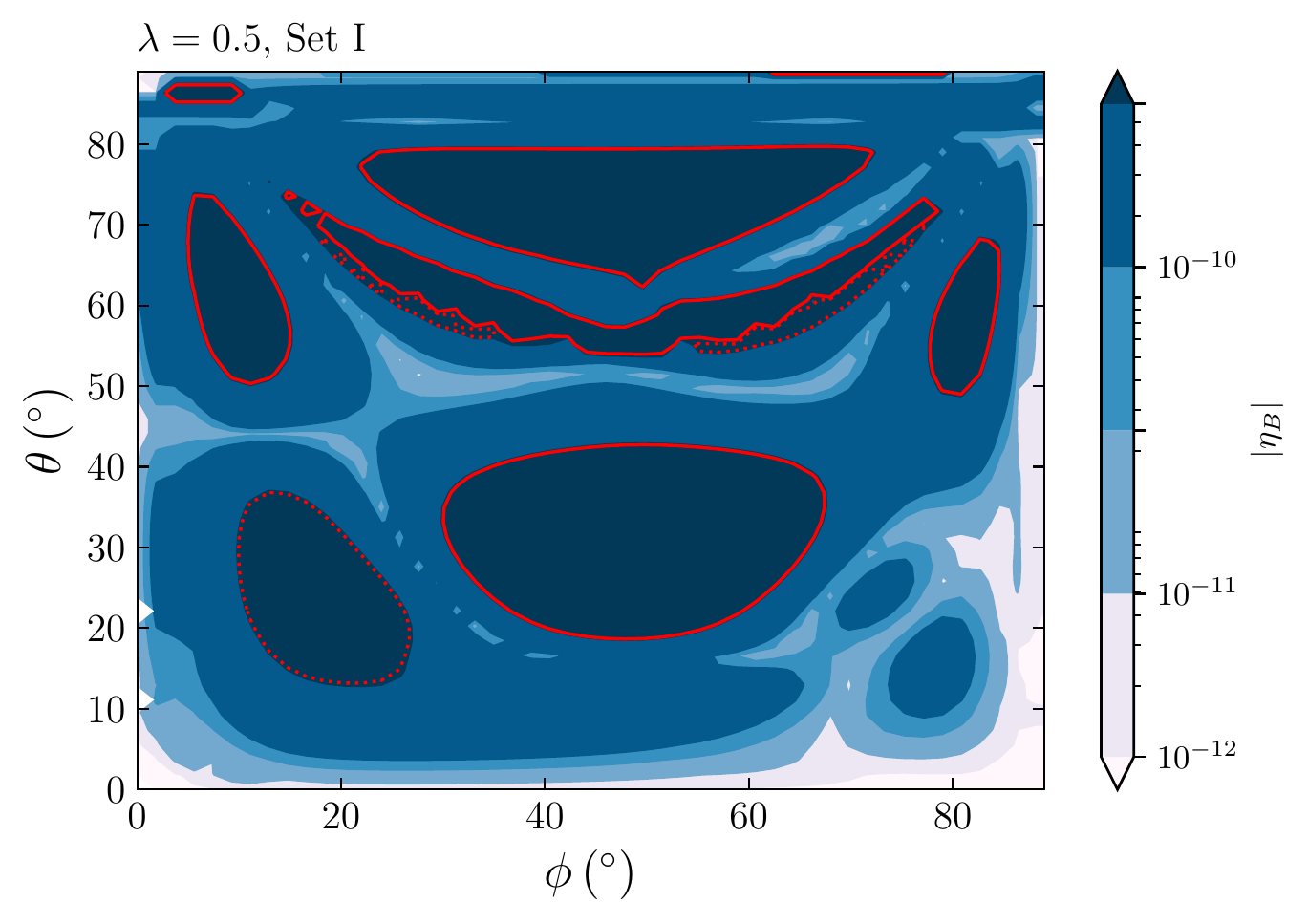}
    \includegraphics[width = .48\textwidth]{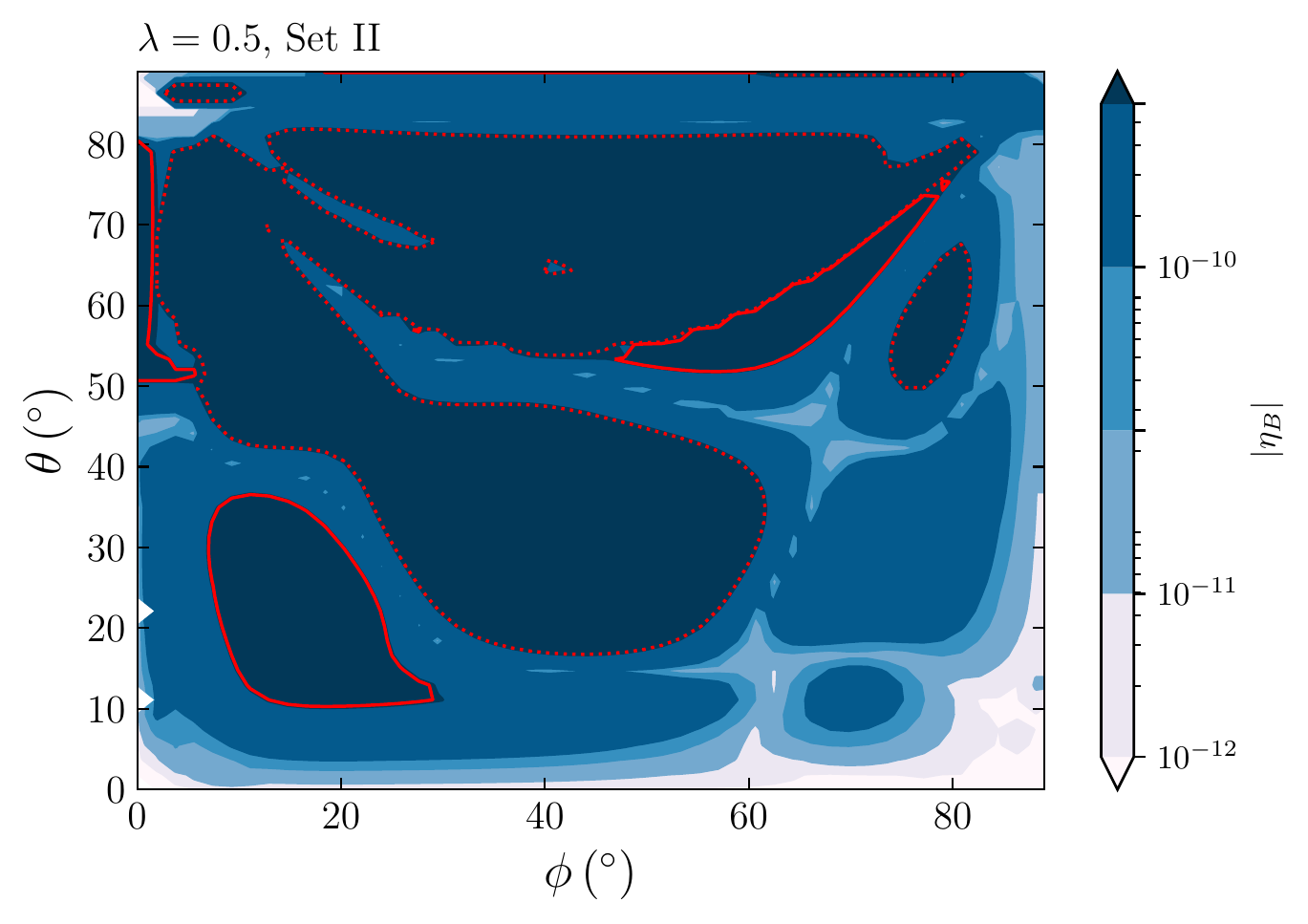}
        \includegraphics[width = .48\textwidth]{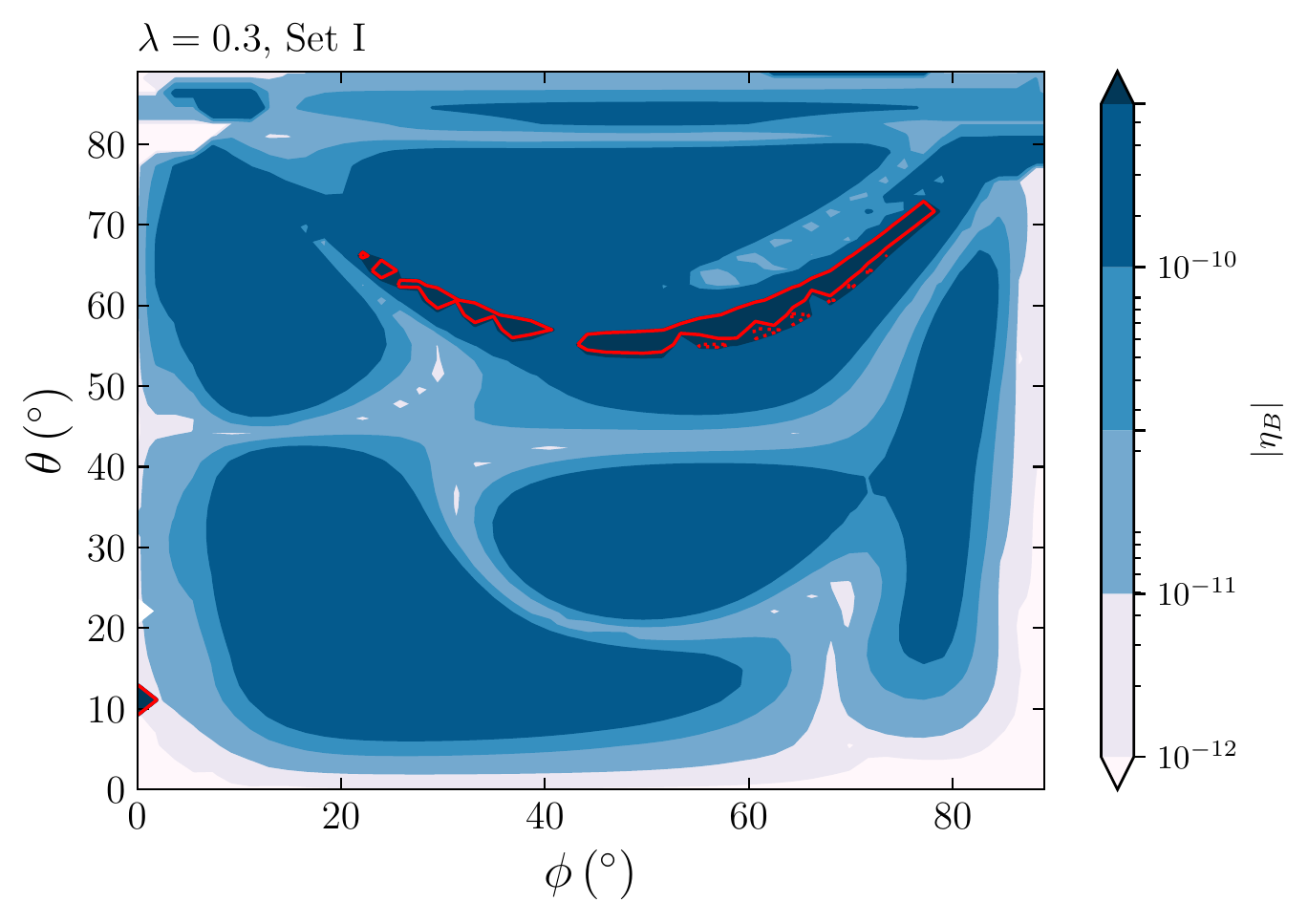}
    \includegraphics[width = .48\textwidth]{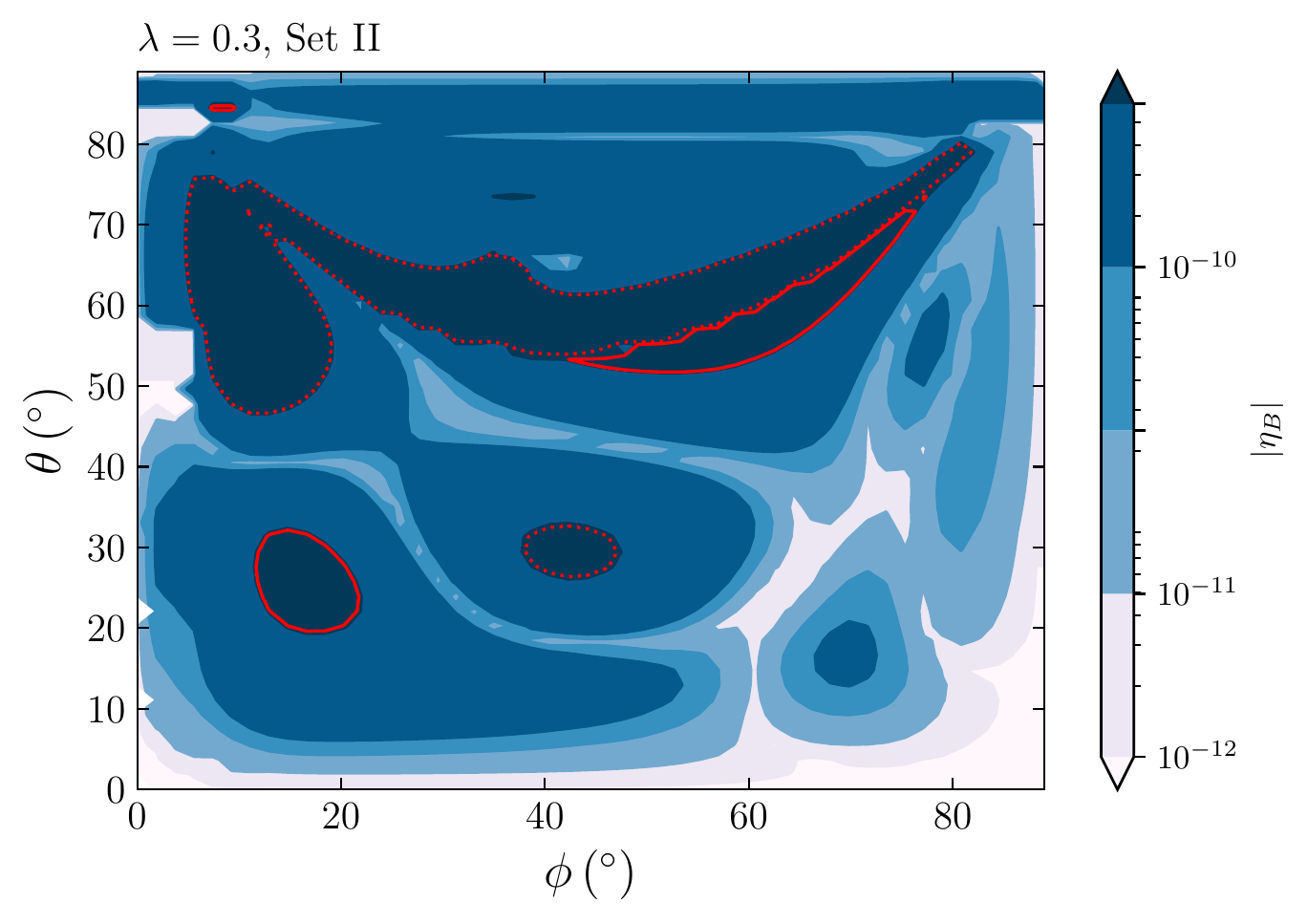}
        \includegraphics[width = .48\textwidth]{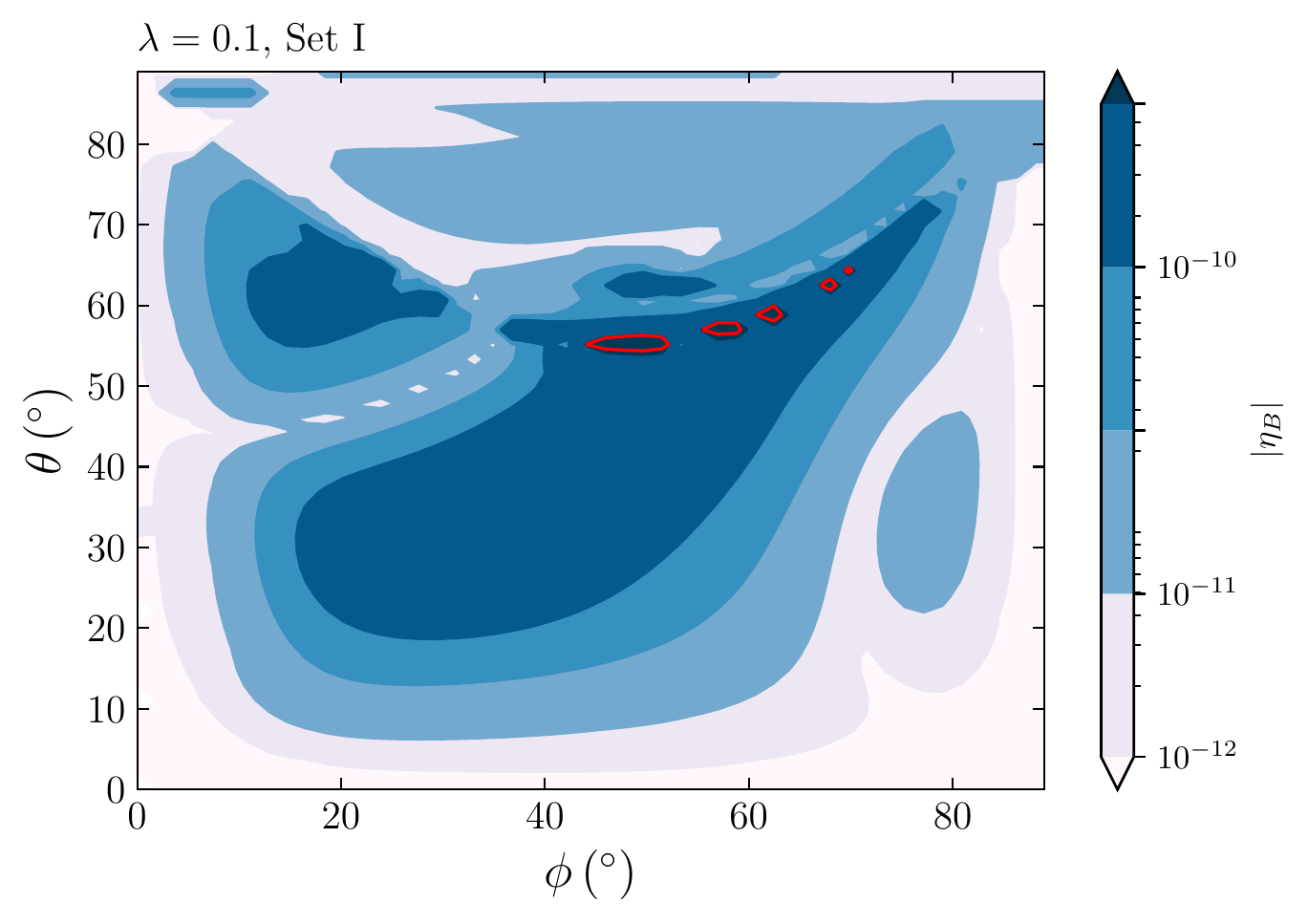}
    \includegraphics[width = .48\textwidth]{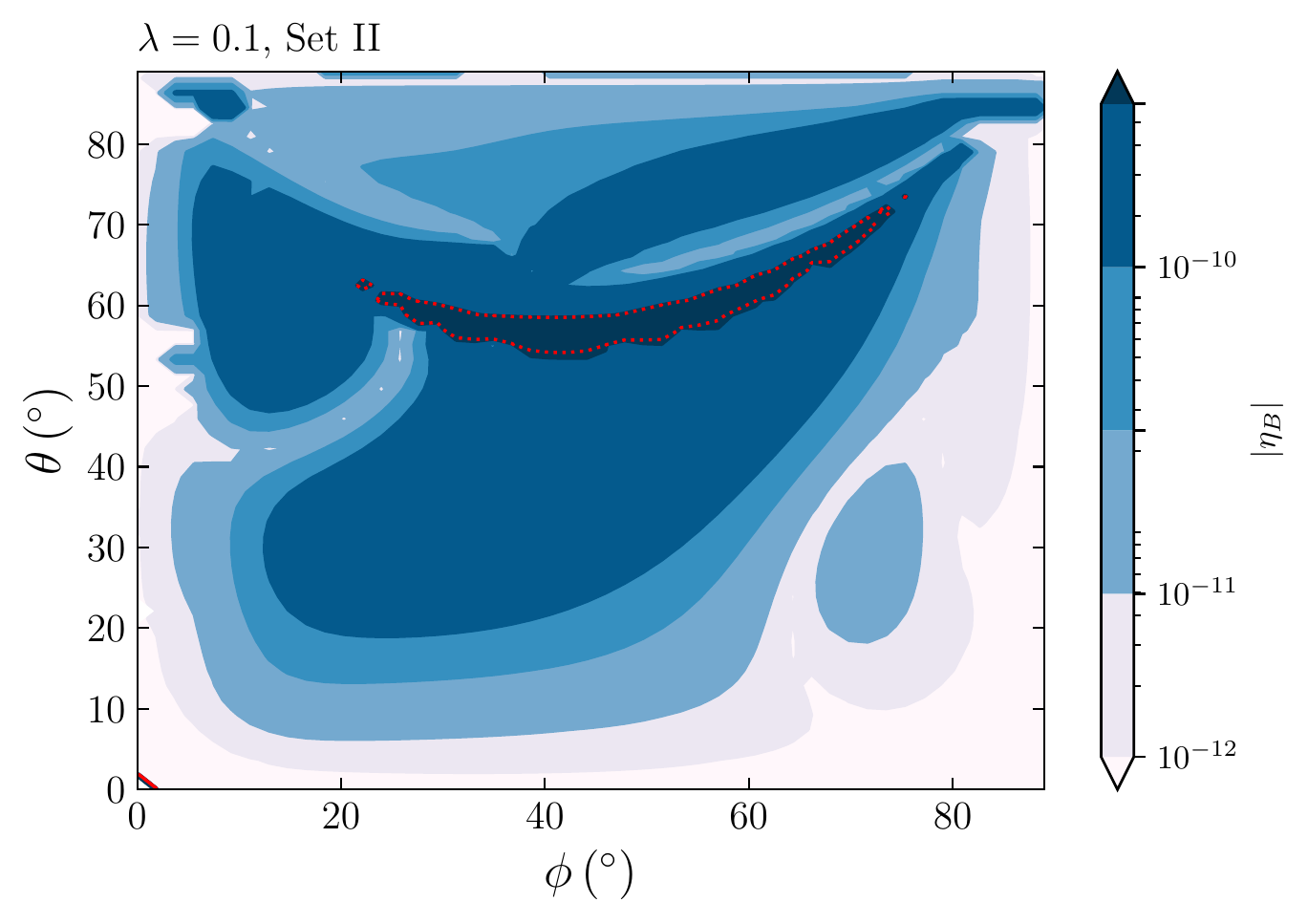}
    \caption{The contour plots in the $\phi-\theta$ plane of the BAU (in modulus) predicted by the DMEs with three decaying right-handed neutrinos in the case of thermal initial abundance. The details of the plots are as in Fig.~\ref{fig:LG_Scan}.}
    \label{fig:LG_Scan_TIA}
\end{figure}

\section{The Impact of Resonance Effects}\label{App:resonance}
In this appendix, we estimate the relevance of the resonance effects of the CP-asymmetry in our numerical analysis. We first note that $\Delta M_{kj}/\Gamma_j \propto \lambda^{-2}$ meaning that, for a given point in the $\phi-\theta$ plane, the resonance effects are depleted if $\lambda$ is sufficiently small. We show in Fig.~\ref{fig:res_scan} the points of our scan for which we find that $\Delta M_{21}/\Gamma_1 \lesssim 10$ (left panel) and $\Delta M_{32}/\Gamma_2 \lesssim 10$ (right panel) for $\lambda = 1$ (top panels) and $0.5$ (bottom panels). The figure is obtained for the neutrino mixing angles and squared mass differences as in Set I. The results remain qualitatively similar for the input parameters in Set II and are thus not shown here. The plots in the top panels show that, when $\lambda = 1$, we have $\Delta M_{21} /\Gamma_1 \lesssim 10$ and $\Delta M_{32}/\Gamma_2 \lesssim 10$ in the region of the parameter space corresponding to $M_2\lesssim 1.5 M_1$ and $M_3 \lesssim 1.5 M_2$, respectively. When $\lambda = 0.5$, these regions are reduced, and it becomes difficult to find points satisfying the two inequalities for $\lambda < 0.5$. This indicates that resonance effects are negligible in most of the parameter space when $\lambda \lesssim 0.5$. By comparing the plots in the left and right panels, it follows that there is no region where the three right-handed neutrinos are simultaneously in the resonant regime, thus justifying the use of the regularisation in Eq.~\eqref{eq:resonance_reg} in the main analysis.

\begin{figure}[t!]
    \centering
    \includegraphics[width = \textwidth]{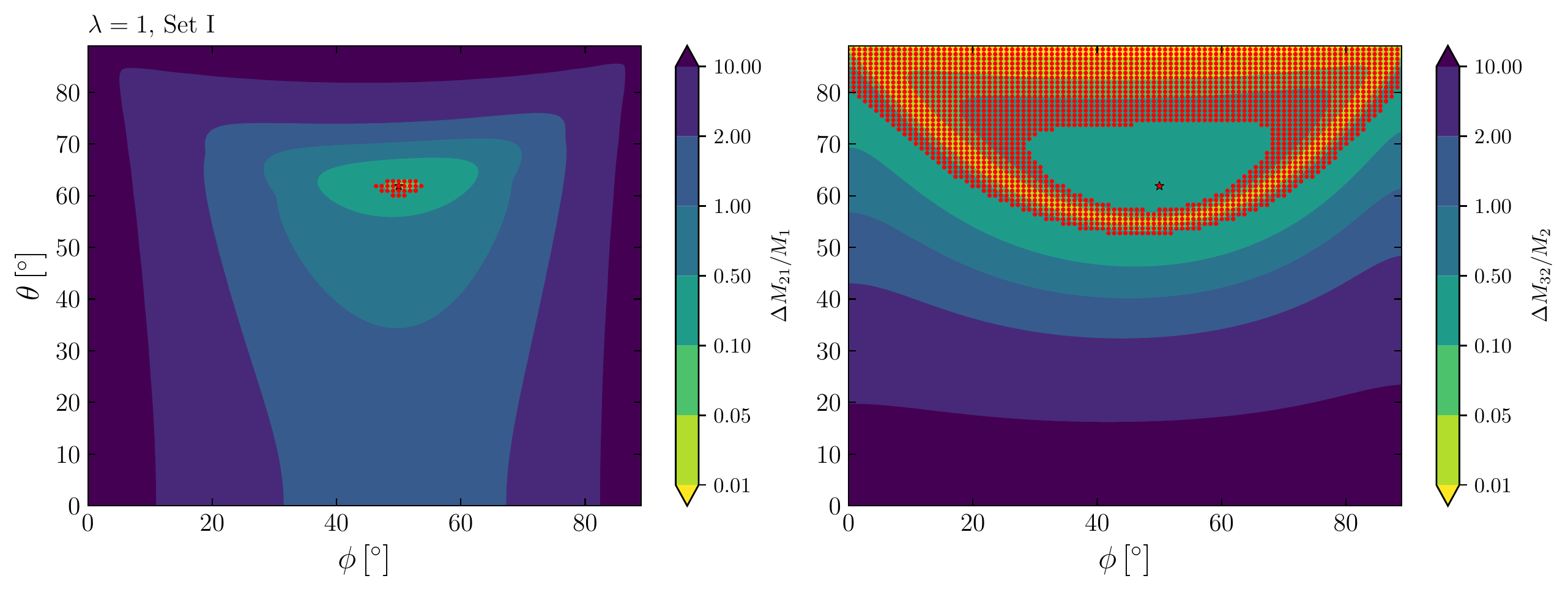}
    \includegraphics[width = \textwidth]{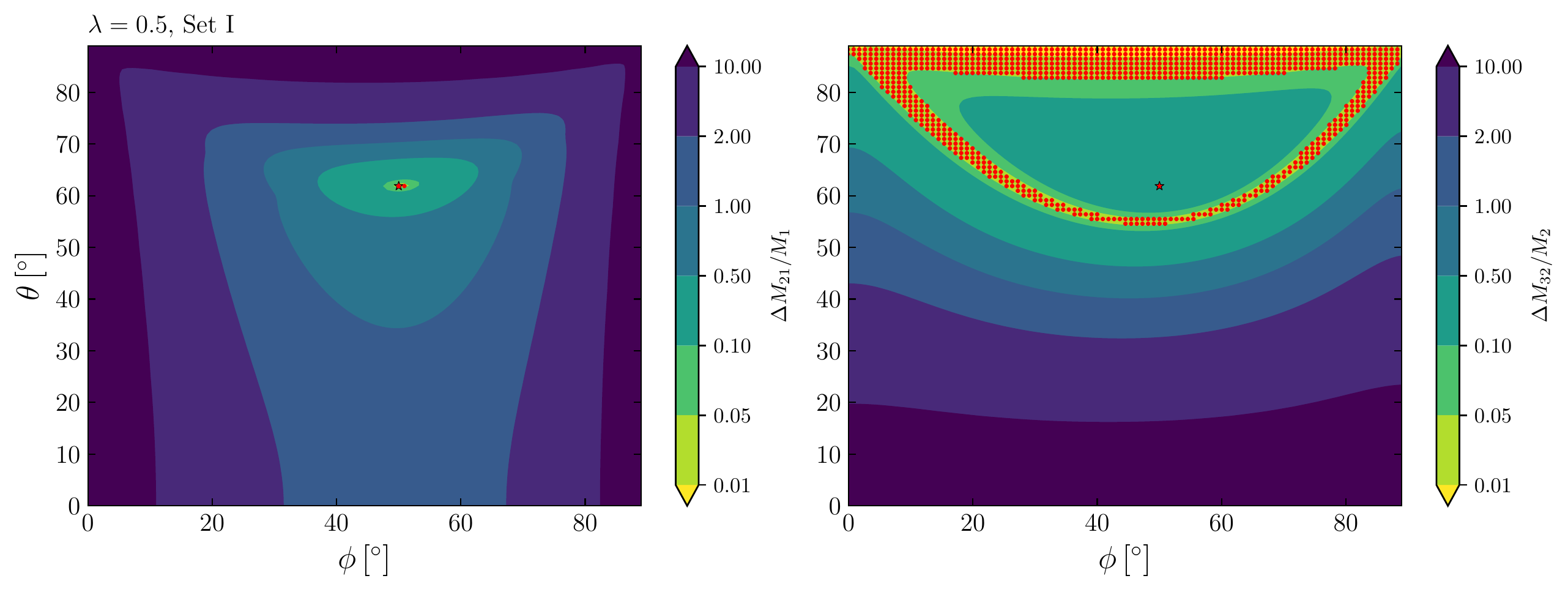}
    \caption{The red points in the left and right panels are those for which we find $\Delta M_{21}/\Gamma_1\lesssim 10$ and $\Delta M_{32}/\Gamma_2\lesssim 10$, respectively. The points in the top (bottom) panels are obtained for $\lambda = 1\,(0.5)$. The figure is obtained for the input parameters as in Set I. The remaining details of the plots are as in the bottom panels of Fig.~\ref{fig:Contour_Mass_ratios_withSK}.}
    \label{fig:res_scan}
\end{figure}

\providecommand{\href}[2]{#2}\begingroup\raggedright\endgroup

\end{document}